\documentclass[a4paper,11pt]{article}
\pdfoutput=1
\usepackage{jheppub}
\usepackage{amsmath,amssymb}
\usepackage[T1]{fontenc}
\usepackage{hyperref}
\usepackage{comment}

\newcommand{\AdS}{\mathrm{AdS}}
\newcommand{\Sph}{\mathrm{S}}
\newcommand{\Tor}{\mathrm{T}}
\newcommand{\str}{\operatorname{str}}
\newcommand{\tr}{\operatorname{tr}}
\newcommand{\Ad}{\operatorname{Ad}}
\newcommand{\fg}{\mathfrak{g}}
\newcommand{\Lax}{\mathcal{L}}
\newcommand{\NFo}{N_{\mathrm{F1}}}
\newcommand{\NDo}{N_{\mathrm{D1}}}
\newcommand{\NDf}{N_{\mathrm{D5}}}
\newcommand{\NNS}{N_{\mathrm{NS5}}}
\newcommand{\QFo}{Q_{\mathrm{F1}}}
\newcommand{\QDo}{Q_{\mathrm{D1}}}
\newcommand{\QDf}{Q_{\mathrm{D5}}}
\newcommand{\QNS}{Q_{\mathrm{NS5}}}

\title{Integrable AdS$_2$ branes in mixed flux AdS$_3$}

\author[a]{Davide Bonomi,}
\author[a,b]{Volker Schomerus\,}
\affiliation[a]{Deutsches Elektronen-Synchrotron DESY,\\
Notkestra{\ss}e 85, 22607 Hamburg, Germany}
\affiliation[b]{II. Institut f\"ur Theoretische Physik,\\ 
Universit\"at Hamburg, 22761 Hamburg, Germany}
\emailAdd{davide.bonomi@desy.de}
\emailAdd{volker.schomerus@desy.de}

\abstract{We show that half-BPS D3-branes wrapping AdS$_2\times$S$^2$ in AdS$_3\times$S$^3\times$T$^4$ with mixed NSNS and RR flux define integrable boundary conditions for the supercoset sigma model of the Green--Schwarz superstring. This extends the integrable D-branes of Dekel and Oz to backgrounds with a Wess--Zumino term, and previous bosonic results for mixed flux to the full superstring. Using the D3-brane action, we find that the electric and magnetic worldvolume fields that stabilise the branes at generic positions are obtained from those of the AdS$_2\times$S$^2$ branes of the WZW model by the same $SL(2,\mathbb{R})$ rotation that maps the pure NSNS background to the mixed flux one, acting on the D3-brane as electric--magnetic duality. On the worldsheet, cancelling the fermionic boundary terms requires introducing a specific closed two-form in the boundary action, and its contribution to the bosonic boundary conditions reproduces precisely these worldvolume fields. We then construct field-dependent reflection matrices that lift the boundary conditions to gluing conditions for the Cagnazzo--Zarembo Lax connection and describe displacements along geodesics normal to the brane. The resulting double-row monodromy matrix generates infinitely many conserved charges, which establishes the classical integrability of the open superstring for any mixture of fluxes. In AdS$_3$/CFT$_2$, these branes are expected to be dual to conformal interfaces in a deformation of the symmetric product orbifold of T$^4$.}

\begin{document}
\maketitle
\flushbottom

\section{Introduction}\label{sec:intro}

String theory in anti-de Sitter (AdS) backgrounds provides one side of the AdS/CFT correspondence~\cite{Maldacena:1997re,Gubser:1998bc,Witten:1998qj}. Soon after the Green--Schwarz superstring on AdS$_5\times$S$^5$ had been formulated as a sigma model on a supercoset~\cite{Metsaev:1998it}, Bena, Polchinski and Roiban discovered that the resulting worldsheet theory is classically integrable~\cite{Bena:2003wd}. This observation has ignited an enormous amount of research on the classical and quantum integrability of strings in AdS$_5\times$S$^5$ and in a number of other backgrounds, see \cite{Arutyunov:2009ga,Beisert:2010jr,Levkovich-Maslyuk:2019awk} for reviews. Backgrounds of the form AdS$_3\times$S$^3\times$T$^4$ play a special role in this context, since they can be supported by any combination of NSNS and RR three-form fluxes. For pure NSNS flux, the worldsheet theory is a WZW model that can be solved exactly~\cite{Giveon:1998ns,Maldacena:2000hw}. For mixed fluxes, Cagnazzo and Zarembo showed that the Green--Schwarz superstring is classically integrable~\cite{Cagnazzo:2012se}, generalising the Lax connection that had been found for pure RR flux~\cite{Chen:2005uj,Babichenko:2009dk}. Beyond the classical level, however, integrability of strings in AdS$_3$ is much less developed than in AdS$_5\times$S$^5$, in particular for mixed fluxes. In spite of some promising steps, most notably the construction of the worldsheet S matrix up to dressing factors~\cite{Hoare:2013pma,Hoare:2013ida,Lloyd:2014bsa} and the study of giant magnons~\cite{Hoare:2013lja}, our understanding of the quantum theory for mixed fluxes is still far from complete. We briefly summarise the current status in section~\ref{sec:conclusions}.

D-branes add an important new element to this story. In the AdS/CFT correspondence, branes that reach the conformal boundary of AdS are dual to interfaces, boundaries or, more generally, defects of higher codimension in the boundary theory. In AdS$_5\times$S$^5$, the prime example is the D5-brane on AdS$_4\times$S$^2$ that was introduced by Karch and Randall~\cite{Karch:2000ct,Karch:2000gx}. It is dual to a codimension-one defect in $\mathcal{N}=4$ super Yang--Mills theory~\cite{DeWolfe:2001pq}. If the brane carries $k$ units of magnetic flux through its S$^2$, the defect becomes an interface between gauge theories with gauge groups $SU(N)$ and $SU(N-k)$. Its field theory description involves a Nahm pole of the type that appears in the classification of half-BPS boundary conditions and interfaces of $\mathcal{N}=4$ super Yang--Mills theory by Gaiotto and Witten~\cite{Gaiotto:2008sa,Gaiotto:2008ak}. Branes with an AdS$_2$ factor, on the other hand, describe line defects such as Wilson lines in higher representations~\cite{Drukker:2005kx}. Integrability in this context was first observed at weak coupling, where the one-loop dilatation operator of the D3--D5 defect theory acts as an integrable open spin chain~\cite{DeWolfe:2004zt}. At strong coupling, classical integrability of open strings was first analysed for the bosonic sigma model~\cite{Mann:2006rh}. Dekel and Oz then constructed integrable boundary conditions for the full Green--Schwarz superstring and showed that they include the Karch--Randall brane~\cite{Dekel:2011ja}. The quantum integrability of the D3--D5 system has since been explored in a series of works, starting with the computation of one-point functions~\cite{deLeeuw:2015hxa,Buhl-Mortensen:2015gfd} and leading to non-perturbative proposals that are based on integrable boundary states~\cite{Komatsu:2020sup,Gombor:2020kgu}. On the string theory side, open strings that end on the D5-brane with flux require a reflection matrix that depends on the embedding coordinates of the string~\cite{Linardopoulos:2021rfq}, see~\cite{deLeeuw:2019usb,Linardopoulos:2025ypq} for reviews.

In AdS$_3\times$S$^3$, the simplest half-BPS branes wrap an AdS$_2\times$S$^2$ subspace. For pure NSNS flux they are well understood from the WZW model on $SL(2,\mathbb{R})\times SU(2)$, in which the S$^2$ and AdS$_2$ factors are localised along conjugacy classes and twisted conjugacy classes, respectively~\cite{Alekseev:1998mc,Bachas:2000fr}. The AdS$_2$ branes reach the boundary of AdS$_3$, and their interpretation as conformal interfaces in the dual two-dimensional conformal field theory was developed in~\cite{Bachas:2001vj}. More recent work on their holographic duals in symmetric product orbifolds is discussed in section~\ref{sec:conclusions}. Much less is known for mixed fluxes, in particular about the integrability of open strings that end on these branes. Integrable boundaries in AdS$_3$ have mostly been studied for pure RR flux~\cite{Prinsloo:2015apa,Bielli:2024xuv}, and for mixed fluxes integrability has so far only been established in the bosonic sector~\cite{Gil:2026ost}.

The present work brings these two lines of research together. A general framework for integrable boundary conditions of Green--Schwarz sigma models on semi-symmetric superspaces was developed by Dekel and Oz~\cite{Dekel:2011ja}. In their approach, the boundary conditions are expressed as gluing conditions for the Lax connection that involve a constant involutive automorphism of the superalgebra, and they found integrable branes in AdS$_5\times$S$^5$ and AdS$_4\times$CP$^3$. AdS$_3$ backgrounds with mixed fluxes are not covered by this analysis, and they differ from these cases in an important way. The Wess--Zumino term that describes the NSNS flux requires a boundary contribution to the action, and gluing conditions with a constant automorphism only describe branes at very special positions~\cite{Gil:2026ost}. In this paper we fill this gap and extend the analysis of integrable branes to the Green--Schwarz superstring on AdS$_3\times$S$^3\times$T$^4$ with mixed NSNS and RR fluxes.

Our starting point is a family of half-BPS D3-branes that wrap AdS$_2\times$S$^2$ and are localised at a point of T$^4$. In the pure NSNS background, their classical stability at generic positions relies on the presence of electric and magnetic worldvolume fields~\cite{Bachas:2000ik,Bachas:2000fr}. We first investigate how these configurations extend to backgrounds with mixed fluxes. Using the Dirac--Born--Infeld and Wess--Zumino actions of the D3-brane, we determine the worldvolume fields for arbitrary positions and values of the NSNS and RR fluxes. The result, see equation~\eqref{eq:rotated-fields}, has a simple interpretation in terms of the electric--magnetic duality of the D3-brane~\cite{Gaillard:1981rj,Gibbons:1995cv,Tseytlin:1996it,Green:1996qg}: the worldvolume fields are obtained from their pure NSNS counterparts by precisely the same duality rotation that relates the two supergravity backgrounds, while the embedding of the brane remains unchanged. This provides a spacetime description of the branes throughout the full range of mixed fluxes, including the pure RR limit. We then derive the corresponding worldsheet boundary conditions for the Green--Schwarz supercoset sigma model. As is familiar from WZW models with boundaries~\cite{Alekseev:1998mc,Gawedzki:2001ye}, the NSNS Wess--Zumino term requires a boundary contribution whose exterior derivative reproduces the pullback of the NSNS three-form to the brane. This leaves the freedom to add a closed two-form, which was exploited in the bosonic analysis of~\cite{Gil:2026ost} to obtain integrable boundary conditions. Different choices of the two-form give consistent boundary conditions in the purely bosonic sigma model, but in the presence of fermions the boundary action must also cancel additional contributions from the Green--Schwarz term. We show that this can be achieved through a specific choice of the closed two-form $\mathcal{F}_0$, given in equations~\eqref{eq:F0a} and~\eqref{eq:F0b}. Remarkably, this form modifies the bosonic boundary conditions so that they reproduce precisely the electric and magnetic worldvolume fields obtained from the spacetime analysis. We thus arrive at the boundary conditions in equation~\eqref{eq:complete-bc}, which incorporate both the bosonic and fermionic degrees of freedom and preserve the half-BPS symmetry of the branes.

Finally, we establish the classical integrability of the resulting open-string theory. Starting from the bosonic analysis of~\cite{Gil:2026ost}, we construct reflection matrices for the full supercoset model, including the fermionic degrees of freedom. The key observation is that the reflection matrices can be built from the currents that generate geodesics normal to the brane, and that they describe finite displacements along these geodesics. We show that the reflection matrices in equation~\eqref{eq:full-reflection} lift the boundary conditions derived from the action to gluing conditions for the complete Lax connection. As for the D5-brane with flux and for the Coulomb branch D3-brane in AdS$_5\times$S$^5$~\cite{Linardopoulos:2021rfq,Demjaha:2025axy}, the reflection matrices are dynamical, i.e.\ they depend on the fields at the boundary. The gluing conditions allow us to construct the double-row monodromy matrix in equation~\eqref{eq:double-row-2}, whose time evolution takes the form of a conjugation, see equation~\eqref{eq:double-row-evolution}. Its spectral invariants therefore generate an infinite family of conserved quantities, which establishes the classical integrability of open strings that end on AdS$_2\times$S$^2$ branes at generic positions, for all values of the fluxes.

The paper is organised as follows. In section~\ref{sec:bulk} we review the mixed flux AdS$_3\times$S$^3\times$T$^4$ background, the corresponding supercoset sigma model and its classical integrability. In section~\ref{sec:branes} we introduce the AdS$_2\times$S$^2$ branes that are the main focus of this paper. We first describe them from the spacetime perspective, and then we construct the open-string worldsheet action and derive the boundary conditions from its variation. In section~\ref{sec:integrability} we prove that these boundary conditions preserve classical integrability by constructing the reflection matrices and the double-row monodromy matrix. In section~\ref{sec:conclusions} we present our conclusions and discuss future directions. Technical details are collected in appendices~\ref{app:potential} and~\ref{app:brane-eom}.

\section{Review of bulk theory and integrability}\label{sec:bulk}

The goal of this section is to review known results on type IIB superstring theory in AdS$_3\times$S$^3\times$T$^4$. We shall start with the spacetime picture in the first subsection before turning to the corresponding worldsheet description in the second. Integrability of the Green--Schwarz superstring on AdS$_3\times$S$^3\times$T$^4$ is finally reviewed in the third subsection, following the construction of Cagnazzo and Zarembo \cite{Cagnazzo:2012se} for mixed NSNS and RR fluxes.

\subsection{F1--NS5--D1--D5 systems and their near-horizon geometry}\label{sec:background}

Our starting point is type IIB string theory on $\mathbb{R}^{1,5}\times\Tor^4$, with coordinates $x^0,\dots,x^5$ along the six non-compact directions and $x^6,\dots,x^9$ along the four-torus. In this spacetime we place $\NFo$ fundamental strings, $\NDo$ D1-branes, $\NDf$ D5-branes and $\NNS$ NS5-branes. All of them extend along the directions $x^0$ and $x^1$, and the two kinds of fivebranes, in addition, wrap the four-torus. In the transverse space $\mathbb{R}^4$, which is spanned by $x^2,\dots,x^5$, all branes sit at the origin. For generic values of the four numbers of branes, the strings and fivebranes form a bound state. Just like the more familiar D1--D5 and F1--NS5 systems, it preserves eight of the thirty-two supercharges of type IIB string theory, i.e.\ one quarter of the supersymmetry, which act as a two-dimensional $\mathcal{N}=(4,4)$ Poincar\'e supersymmetry on the common worldvolume of the branes, see e.g. \cite{David:2002wn} for a review and further references.

The supergravity solution that describes the bound state depends on the numbers of branes and on the asymptotic values of the moduli of type IIB string theory on $\Tor^4$. In the solution, the brane numbers $N_X$ enter through the usual prefactors of the form
\begin{equation}\label{eq:harmonic-functions}
f_X = 1 + \frac{Q_X}{r^2} = 1 + \frac{2\pi \alpha'^2 g_6^2 \mathcal{T}_X}{r^2}\,,\qquad g_6^2=\frac{g_s^2}{v}\,,
\end{equation}
where $r$ denotes the distance from the branes in the four-dimensional transverse space, $\mathcal{T}_X$ are the tensions of the branes along the two-dimensional plane that is spanned by $x^0$ and $x^1$ and $g_6$ is the six-dimensional string coupling, with $v$ denoting the volume of the torus in units of $(2\pi\sqrt{\alpha'})^4$, see e.g.\ \cite{Tseytlin:1996as,Maldacena:1997re,Giveon:1998ns,David:2002wn}. Inserting standard expressions for the tensions of the four types of branes one finds that the charge parameters $Q_X$ are given by
\begin{equation}\label{eq:charge-parameters}
\QNS=\alpha'\NNS\,,\qquad \QDf=\alpha' g_s\NDf\,,\qquad \QFo=\frac{\alpha' g_s^2\NFo}{v}\,,\qquad \QDo=\frac{\alpha' g_s\NDo}{v}\,.
\end{equation}
In order to display the relevant supergravity solution we shall make the simplifying assumption that the charge parameters of the strings and of the fivebranes agree pairwise,
\begin{equation}\label{eq:attractor}
\QFo=\QNS\,,\qquad \QDo=\QDf\,.
\end{equation}
This choice implies that both the dilaton $\Phi$ and the volume $v$ of the four-dimensional torus remain constant, i.e.\ are independent of $r$. Their values are then given in terms of the numbers of branes by
\begin{equation}\label{eq:attractor-moduli}
v=\frac{\NDo}{\NDf}\,,\qquad g_6^2\equiv\frac{g_s^2}{v}=\frac{\NNS}{\NFo}\,.
\end{equation}
For more general choices of the charge parameters, the scalar fields vary along the radial direction, and in general the RR axion $C_0$ and the component of $C_4$ along the torus are sourced as well. Since the near-horizon background enters the worldsheet theory only through the parameters $\lambda$ and $\chi$, see the comments at the end of this subsection, the condition~\eqref{eq:attractor} is no loss of generality for our purposes. For this choice, the supergravity solution involves a single harmonic function $f$. In string frame, it reads
\begin{equation}\label{eq:sugra}
\begin{aligned}
ds^2&=f^{-1}\big(-(dx^0)^2+(dx^1)^2\big)+f\,\big(dr^2+r^2ds^2_{\Sph^3}\big)
+ds^2_{\Tor^4}\,,\qquad e^{\Phi}=g_s\,,\\[2pt]
\text{H}&=2\QNS\,\mathrm{vol}_{\Sph^3}+\frac{\QFo}{Q}\,df^{-1}\wedge dx^0\wedge dx^1\,,\\[2pt]
g_sF_3&=2\QDf\,\mathrm{vol}_{\Sph^3}+\frac{\QDo}{Q}\,df^{-1}\wedge dx^0\wedge dx^1\,,\\[2pt]
f&=1+\frac{Q}{r^2}\,,\qquad Q=\sqrt{\QNS^2+\QDf^2}=\sqrt{\QFo^2+\QDo^2}\,.
\end{aligned}
\end{equation}
Here $ds^2_{\Sph^3}$ and $\mathrm{vol}_{\Sph^3}$ are the metric and the volume form of the unit three-sphere, and $ds^2_{\Tor^4}$ is a flat metric on the torus with volume $(2\pi\sqrt{\alpha'})^4v$. The magnetic parts of the three-forms are sourced by the fivebranes and the electric parts by the strings. Indeed, the numbers of NS5-branes and D5-branes are measured by the fluxes of $\text{H}$ and $F_3$ through S$^3$, in units of $4\pi^2\alpha'$, while the numbers of fundamental strings and D1-branes are measured by the fluxes of the Hodge duals $e^{-2\Phi}{*\text{H}}$ and ${*F_3}$ through S$^3\times$T$^4$, in units of $(2\pi)^6\alpha'^3$. With the charge parameters in equation~\eqref{eq:charge-parameters}, the solution reproduces all four numbers of branes. It is obtained from an F1--NS5 solution with a single harmonic function by a rotation of the doublet $(\text{H},g_sF_3)$, which is part of the $SL(2,\mathbb{R})$ symmetry of type IIB supergravity and leaves the metric and the dilaton invariant, see e.g.~\cite{Schwarz:1995dk,Hoare:2013pma,Kluson:2016dca} and references therein. This rotation will reappear in subsection~\ref{sec:spacetime-branes}.

The near-horizon geometry is obtained through the decoupling limit of Maldacena \cite{Maldacena:1997re}, which sends $\alpha'$ to zero while keeping
\begin{equation}\label{eq:U}
U=\frac{r}{\alpha'}
\end{equation}
fixed, along with $g_s$, $v$ and the numbers of branes. In units of $\alpha'$, the background then approaches AdS$_3\times$S$^3\times$T$^4$, with
\begin{equation}\label{eq:nh-background}
\begin{aligned}
ds^2&=R^2\big(ds^2_{\AdS_3}+ds^2_{\Sph^3}\big)+ds^2_{\Tor^4}\,,\\[2mm]
\text{H}&=2R^2\chi\,\big(\mathrm{vol}_{\AdS_3}+\mathrm{vol}_{\Sph^3}\big)\,,\qquad e^{\Phi}F_3=2R^2\kappa\,\big(\mathrm{vol}_{\AdS_3}+\mathrm{vol}_{\Sph^3}\big)\,,
\end{aligned}
\end{equation}
where $ds^2_{\AdS_3}$ and $\mathrm{vol}_{\AdS_3}$ are the metric and the volume form of unit-radius AdS$_3$. The space AdS$_3$ is formed by the directions $x^0$ and $x^1$ together with the radial coordinate, and S$^3$ is the sphere that surrounds the branes in $\mathbb{R}^4$. Both have the same radius $R$, while the dilaton and the torus keep their constant values. Comparing with the solution in equation~\eqref{eq:sugra}, we read off the radius and the two coefficients,
\begin{equation}\label{eq:lambda}
\lambda\equiv\frac{R^4}{\alpha'^2}= \NNS^2+g_s^2\NDf^2=g^2_6 N\,,\qquad
\chi^2=\frac{\NFo\NNS}{N}\,,\qquad \kappa^2=\frac{\NDo\NDf}{N}\,,
\end{equation}
where the integer $N$ pairs the charges of the strings with those of the fivebranes,
\begin{equation}\label{eq:N}
N=\NDo\NDf+\NFo\NNS\,.
\end{equation}
As a consequence, the coefficients $\kappa$ and $\chi$ obey the following simple relation
\begin{equation}\label{eq:chi-kappa}
\chi^2+\kappa^2=1\,,
\end{equation}
which the supergravity equations of motion require for any background of the form given in equation~\eqref{eq:nh-background}.

Under the $SL(2,\mathbb{R})$ duality group of type IIB supergravity, the charges $(\NFo,\NDo)$ of the strings and $(\NNS,\NDf)$ of the fivebranes transform as a doublet and its dual, and $N$ is their invariant pairing. According to equation~\eqref{eq:lambda}, $\chi^2$ is the share of $N$ that is carried by the fundamental strings and NS5-branes. Moreover, comparing the NSNS fluxes in equations~\eqref{eq:sugra} and~\eqref{eq:nh-background}, we find
\begin{equation}\label{eq:chi-Q}
\chi\,R^2=\QNS\,.
\end{equation}
Together with the first relations in equations~\eqref{eq:charge-parameters} and~\eqref{eq:lambda}, this shows that 
\begin{equation}\label{eq:flux-quantisation}
\chi\sqrt\lambda=\NNS\,.
\end{equation}
For other asymptotic values of the moduli, the RR axion and the component of $C_4$ along the torus generally take nonzero values in the near-horizon region, where they modify the relations between the fluxes and the numbers of fundamental strings, D1-branes and D5-branes. The relation in equation~\eqref{eq:flux-quantisation} and the expression $g_6\sqrt N$ for $\sqrt\lambda$ remain valid, with $g_6$ evaluated in the near-horizon region, but $\lambda$ and $\chi$ are no longer fixed by the numbers of branes and vary continuously with the moduli \cite{OhlssonSax:2018hgc}.

The two endpoints of the range of $\chi$ are the D1--D5 system, with pure RR flux, and the F1--NS5 system, with pure NSNS flux, which are related by S-duality. For these systems, one of the conditions in equation~\eqref{eq:attractor} is empty, so that $g_6$ and $v$, respectively, remain free. In the F1--NS5 case, the worldsheet theory famously involves the WZW model on $SL(2,\mathbb{R})\times SU(2)$ at level $\NNS$ \cite{Giveon:1998ns,Maldacena:2000hw}.
\medskip

Let us now finish this subsection with a few more general comments on the expected worldsheet description of strings in AdS$_3\times$S$^3$. From the preceding discussion, the worldsheet model must depend on the string tension $\sqrt\lambda$ and on the fraction $\chi$ of NSNS flux. The geometry of the torus only enters the dynamics along T$^4$. The classical worldsheet theory, which we study in this paper, describes the regime of large $\lambda$, and string interactions, which are controlled by $g_6$, are suppressed if $N$ is large compared to $\lambda$. Since the string tension appears as an overall factor of the classical action, the quantisation of $\chi\sqrt\lambda$ plays no role at this level, and we treat $\chi$ as a continuous parameter.

In the near-horizon limit the supersymmetry is enhanced. The bosonic isometries of AdS$_3\times$S$^3$ form the group $SO(2,2)\times SO(4)$, which is locally isomorphic to two copies of $SL(2,\mathbb{R})\times SU(2)$, and the number of supercharges doubles from eight to sixteen. Isometries and supercharges together generate the superisometry group
\begin{equation}\label{eq:superisometry}
PSU(1,1|2)\times PSU(1,1|2)\,.
\end{equation}
Each factor contains one copy of $SU(1,1)\cong SL(2,\mathbb{R})$ and one copy of $SU(2)$, along with eight of the sixteen supercharges. The torus contributes the translations $U(1)^4$ and it does not affect the number of supercharges, since the Killing spinors of the background are constant along T$^4$. The space AdS$_3\times$S$^3$ is the group manifold of $SU(1,1)\times SU(2)$, i.e.\ the quotient of two copies of this group by its diagonal subgroup. Extending the quotient by one fermionic direction for each of the sixteen supercharges leads to the supercoset
\begin{equation}\label{eq:supercoset}
\mathcal{M}=\frac{PSU(1,1|2)\times PSU(1,1|2)}{SU(1,1)\times SU(2)}\,,
\end{equation}
so that the near-horizon geometry is the bosonic body of $\mathcal{M}\times\Tor^4$. The supercoset $\mathcal{M}$ is the target space of the worldsheet theory in subsection~\ref{sec:coset}. Note that it contains only sixteen fermionic directions, whereas the Green--Schwarz superstring has thirty-two fermionic coordinates. We will come back to the precise relation at the beginning of
the next subsection.

\subsection{Green--Schwarz sigma model on the supercoset}\label{sec:coset}

The worldsheet theory of the superstring in the near-horizon background is described by the Green--Schwarz (GS) action. Its fermionic coordinates form two Majorana--Weyl spinors of sixteen real components each, i.e.\ there is one fermionic coordinate for each of the thirty-two supercharges of type IIB string theory. Half of them can be removed by fixing the kappa symmetry of the GS action. In the AdS$_3\times$S$^3\times$T$^4$ background, the thirty-two fermions split into sixteen that are associated with the preserved supercharges and sixteen that are associated with broken ones. Together with the bosonic coordinates of AdS$_3\times$S$^3$, the former parametrise the supercoset $\mathcal{M}$ in equation~\eqref{eq:supercoset}. The GS action restricted to these fields is a sigma model on $\mathcal{M}$ \cite{Pesando:1998wm,Rahmfeld:1998zn,Park:1998un,Metsaev:2000mv}, which possesses its own kappa symmetry of rank eight \cite{Babichenko:2009dk,Cagnazzo:2012se}. It thus describes eight of the sixteen physical fermions, namely those that are massive in light-cone gauge. The remaining eight physical fermions reside outside the coset and are massless. As argued in reference \cite{Babichenko:2009dk}, the classical GS action can be consistently truncated to the coset fields after partial fixing of the kappa symmetry, and in this truncation the bosons along the torus decouple as free fields. In this paper we restrict to the supercoset sigma model. 
The fermions outside the coset were included in the GS action to quadratic order in references~\cite{Sundin:2012gc,Sundin:2013ypa}, and the corresponding massless modes enter the exact worldsheet S matrix~\cite{Borsato:2014hja}.

The supercoset $\mathcal{M}$ is an example of a semi-symmetric superspace, i.e.\ a quotient of a supergroup by a bosonic subgroup that is left invariant by a $\mathbb{Z}_4$ automorphism \cite{Serganova:1983,Berkovits:1999zq,Zarembo:2010sg}. Sigma models on such spaces generalise the construction of Metsaev and Tseytlin for AdS$_5\times$S$^5$ \cite{Metsaev:1998it}. More specifically, $\mathcal{M}$ belongs to the class of permutation cosets $G\times G/G^B_{\mathrm{diag}}$, where $G^B$ denotes the bosonic subgroup of a supergroup $G$ \cite{Babichenko:2009dk,Cagnazzo:2012se}. As we saw in the previous subsection, the bosonic body of such a coset is the group manifold of $G^B$. Cagnazzo and Zarembo observed that this property makes it possible to add a Wess--Zumino term that describes the NSNS flux \cite{Cagnazzo:2012se}. In the remainder of this subsection we review their construction and set up the conventions that we use throughout the rest of the paper.
\medskip

Before we can spell out the action, we actually need some time to discuss the fields that enter the model. Let $G=PSU(1,1|2)$ and let $G^B$ denote its bosonic subgroup $SU(1,1)\times SU(2)$. For matrix computations we use two realisations \(G_L\) and \(G_R\), defined by
\begin{equation}\label{eq:real}
    g_L^\dagger H_Lg_L=H_L,\qquad g_R^\dagger H_Rg_R=H_R,\qquad H_L=\operatorname{diag}(-\sigma_2,\mathbf 1_2),\qquad H_R=\operatorname{diag}(\sigma_2,\mathbf 1_2),
\end{equation}
with \(g_L\in G_L\) and \(g_R\in G_R\). Here the dagger denotes the usual Hermitian conjugation of Grassmann-valued matrices, with conjugation reversing products. The two realisations have the same bosonic subgroup and opposite reality conditions on their odd blocks. Matrix products are understood in the common complexification. The upper-left and lower-right blocks of matrices correspond to the bosonic factors \(SU(1,1)\) and \(SU(2)\), respectively. We label these blocks by \(A\) and \(S\) and use the supertrace
\begin{equation}\label{eq:str}
\str X=\tr X_A-\tr X_S\,.
\end{equation}
The fields of the sigma model we are about to spell out are pairs $(g_L,g_R)$ of $G$-valued functions of the worldsheet coordinates $\sigma$ and $\tau$. They are subject to gauge transformations and global symmetry transformations that act as
\begin{equation}\label{eq:gauge-global}
(g_L,g_R)\longmapsto(h_L\,g_L\,h,\;h_R\,g_R\,h)\,, \qquad h = h(\sigma,\tau)\in G^B\,,\quad h_L,h_R\in G\,.
\end{equation}
The gauge transformation acts on both factors from the right with the same bosonic element $h$, while the global $G\times G$ symmetry with (constant) elements $h_L$ and $h_R$ acts on the two factors independently from the left. Since we are ultimately interested in open strings, we take the worldsheet $M$ to be a strip with coordinates $\tau\in\mathbb{R}$ and $\sigma\in[0,\pi]$. Our conventions for the volume form and the Hodge star are
\begin{equation}\label{eq:orientation}
\mathrm{vol}_M=d\tau\wedge d\sigma\,,\qquad *d\tau=d\sigma\,,\qquad *d\sigma=d\tau\,.
\end{equation}
The action and the equations of motion are built from the left-invariant currents and their decomposition into even and odd parts,\footnote{The labels $L,R$ refer to the two factors of $G\times G$, not to left- or right-invariant currents.}
\begin{equation}\label{eq:currents}
j_{L,R}=g_{L,R}^{-1}dg_{L,R}=j^\tau_{L,R}\,d\tau+j^\sigma_{L,R}\,d\sigma\,,
\qquad
j_{L,R}=j^B_{L,R}+j^F_{L,R}\,.
\end{equation}
The even (bosonic) parts $j^B_{L,R}$ are block diagonal and the odd (fermionic) parts $j^F_{L,R}$ are block off-diagonal. The currents are invariant under the global symmetry and satisfy the Maurer--Cartan equations
\begin{equation}\label{eq:MC-LR}
dj_{L,R}+j_{L,R}\wedge j_{L,R}=0\,.
\end{equation}
Under the gauge transformations in equation~\eqref{eq:gauge-global}, on the other hand, the even and odd parts of the currents transform as
\begin{equation}\label{eq:gauge-currents}
j^B_{L,R}\longmapsto h^{-1}j^B_{L,R}\,h+h^{-1}dh\,,\qquad j^F_{L,R}\longmapsto h^{-1}j^F_{L,R}\,h\,.
\end{equation}
In particular, the inhomogeneous term $h^{-1}dh$ is the same for both factors, so that the difference of the two even currents transforms covariantly. At this point we have built four sets of currents that are labelled by $L/R$ and $B/F$. It turns out that it is advantageous to work with four specific linear combinations instead, which diagonalise the $\mathbb{Z}_4$ action we briefly mentioned before.
\smallskip

Indeed, the (complexified) superalgebra $\fg\oplus\fg$ of $G\times G$ possesses a $\mathbb{Z}_4$ automorphism that exchanges the two summands, with an additional sign on odd elements \cite{Babichenko:2009dk,Cagnazzo:2012se}. Its invariant subalgebra is the diagonal bosonic subalgebra, i.e.\ the Lie algebra of the gauge group. Decomposing the currents into eigencomponents of this automorphism leads to the four $\fg$-valued one-forms\footnote{In \cite{Cagnazzo:2012se}, the $\fg$-valued currents $j_n$ are denoted by $J_n$.}
\begin{equation}\label{eq:Z4-components}
\begin{aligned}
j_0&=\tfrac12\big(j^B_L+j^B_R\big)\,,&\qquad
j_1&=\tfrac12\big(j^F_L+i\,j^F_R\big)\,,\\
j_2&=\tfrac12\big(j^B_L-j^B_R\big)\,,&\qquad
j_3&=\tfrac12\big(j^F_L-i\,j^F_R\big)\,,
\end{aligned}
\end{equation}
where the subscript indicates the $\mathbb{Z}_4$ charge. Conversely, the even and odd parts of the original currents are recovered from
\begin{equation}\label{eq:Z4-inverse}
j^B_L=j_0+j_2\,,\qquad j^B_R=j_0-j_2\,,\qquad
j^F_L=j_1+j_3\,,\qquad j^F_R=-i\,(j_1-j_3)\,.
\end{equation}
According to equation~\eqref{eq:gauge-currents}, the component $j_0$ transforms as a connection, while the remaining components transform in the adjoint representation of the gauge group,
\begin{equation}\label{eq:gauge-Z4}
j_0\longmapsto h^{-1}j_0\,h+h^{-1}dh\,,\qquad
j_n\longmapsto h^{-1}j_n\,h\,,\quad n=1,2,3\,.
\end{equation}
Hence $j_0$ plays the role of a non-dynamical gauge field for the coset symmetry. The bosonic degrees of freedom of the string are carried by $j_2$, and the fermionic ones by $j_1$ and $j_3$.

The Maurer--Cartan equations~\eqref{eq:MC-LR} can be brought into a manifestly gauge-covariant form \cite{Berkovits:1999zq,Cagnazzo:2012se}. To this end we introduce the field strength of $j_0$ and the covariant exterior derivative of a $\fg$-valued $p$-form $C$,
\begin{equation}\label{eq:F-D}
F=dj_0+j_0\wedge j_0\,,\qquad DC=dC+j_0\wedge C+(-1)^{p+1}\,C\wedge j_0\,.
\end{equation}
Taking linear combinations of the even and odd parts of the Maurer--Cartan equations for $j_L$ and $j_R$, one finds
\begin{equation}\label{eq:MC-Z4}
\begin{aligned}
F+j_2\wedge j_2+j_1\wedge j_3+j_3\wedge j_1&=0\,,\\
Dj_2+j_1\wedge j_1+j_3\wedge j_3&=0\,,\\
Dj_1+j_2\wedge j_3+j_3\wedge j_2&=0\,,\\
Dj_3+j_2\wedge j_1+j_1\wedge j_2&=0\,.
\end{aligned}
\end{equation}
These relations are kinematic identities that hold independently of the dynamics. Below they are supplemented by the equations of motion of the mixed flux sigma model.
\medskip

After all this lengthy preparation we are finally able to introduce the relevant sigma model and discuss its properties. On a closed worldsheet $M$, the action of the mixed flux sigma model consists of three terms \cite{Cagnazzo:2012se}\footnote{We denote the physical NSNS field strength by $\text{H}$ and
reserve $H$ for the dimensionless superspace three-form in the sigma model. On the bosonic body,
\begin{equation*}
\text{H}=R^2\chi\,H\big|_{\mathrm{bos}}.
\end{equation*} },
\begin{equation}\label{eq:action}
S=\frac{\sqrt\lambda}{2\pi}\int_M\str\big(j_2\wedge*j_2+\kappa\,j_1\wedge j_3\big) +\frac{\sqrt\lambda}{2\pi}\,\chi\int_{\mathcal{B}}H\,.
\end{equation}
The first term is the kinetic term of the bosonic fields. The second term is the GS term, which couples the fermions to the RR flux \cite{Metsaev:1998it,Berkovits:1999zq}. The last term is the Wess--Zumino term, which describes the NSNS flux. It involves a three-manifold $\mathcal{B}$ with boundary $M$, to which the fields are extended, and the three-form\footnote{It is also possible to write the action in terms of $\mathfrak{g}\oplus\mathfrak{g}$-valued currents. In this formulation, the Wess--Zumino term involves the difference of the supertraces in the two factors \cite{Hoare:2013pma}.}
\begin{equation}\label{eq:H}
H=2\str\Big(\tfrac23\,j_2\wedge j_2\wedge j_2 +j_1\wedge j_3\wedge j_2+j_3\wedge j_1\wedge j_2\Big)\,.
\end{equation}
Its purely bosonic part is the standard Wess--Zumino three-form on the group manifold $SU(1,1)\times SU(2)$. The fermionic terms are determined, up to exact forms, by the requirement that the variation of the Wess--Zumino term is local, i.e.\ that it can be written as an integral over $M$ \cite{Cagnazzo:2012se}. The three-form $H$ is closed as a consequence of the Maurer--Cartan equations~\eqref{eq:MC-Z4}.

The parameters of the action~\eqref{eq:action} are identified with those of the near-horizon geometry by considering purely bosonic configurations. In this case we may use the gauge symmetry~\eqref{eq:gauge-global} to set $g_R=1$, so that
\begin{equation}\label{eq:bosonic-gauge}
j_0=j_2=\tfrac12\,g^{-1}dg\,,\qquad g=g_L\,,
\end{equation}
and the action becomes
\begin{equation}\label{eq:action-bos}
S_{\mathrm{bos}}=\frac{\sqrt\lambda}{8\pi}\int_M \str\big(g^{-1}dg\wedge*g^{-1}dg\big) +\frac{\chi\sqrt\lambda}{12\pi}\int_{\mathcal{B}}\str\big(g^{-1}dg\wedge g^{-1}dg\wedge g^{-1}dg\big)\,.
\end{equation}
This is the principal chiral model on $SU(1,1)\times SU(2)$ with a Wess--Zumino term. Its target space is AdS$_3\times$S$^3$ with radius $R$, where $R^2/\alpha'$ is given by $\sqrt\lambda$ in agreement with equation~\eqref{eq:lambda}, and the coefficient of its Wess--Zumino term is proportional to $\chi\sqrt\lambda$. For $\chi=1$ the action~\eqref{eq:action-bos} is that of the WZW model at level $\sqrt\lambda$. For the path integral to be independent of the extension of the fields to $\mathcal{B}$, the coefficient $\chi\sqrt\lambda$ must be an integer \cite{Witten:1983ar}, in agreement with the flux quantisation in equation~\eqref{eq:flux-quantisation}. Similarly, the fermionic GS term that appears in the action~\eqref{eq:action} is proportional to the RR flux parameter $\kappa$. In the near-horizon background, the coefficients $\kappa$ and $\chi$ are related by equation~\eqref{eq:chi-kappa}. Cagnazzo and Zarembo showed that the same relation is required for kappa symmetry and one-loop conformal invariance of the sigma model \cite{Cagnazzo:2012se}. So we shall continue to assume that equation~\eqref{eq:chi-kappa} holds. Let us also note that the action requires additional data when the worldsheet has a boundary. We postpone this discussion to section~\ref{sec:branes}.
\smallskip

Now that we have set up the action~\eqref{eq:action}, we can use it to derive the corresponding equations of motion for the involved fields. In order to do so, we
consider variations of the fundamental fields which we write in the form
\begin{equation}\label{eq:variations}
\delta g_{L,R}=g_{L,R}\,\Delta_{L,R}\,, \qquad \delta j_{L,R} = d\Delta_{L,R}+[j_{L,R},\Delta_{L,R}].
\end{equation}
Of course the variations $\Delta_{L,R}$ can also be decomposed into bosonic and fermionic components and we can pass to the variations $\Delta_n$ that diagonalise the $\mathbb{Z}_4$ action much in the same way as we did for the currents in equation~\eqref{eq:Z4-components}. The component $\Delta_0$ generates a gauge transformation, under which the action is invariant. The variations with respect to the remaining components give
\cite{Cagnazzo:2012se}
\begin{equation}\label{eq:eom}
\begin{aligned}
D*j_2-\kappa\,j_1\wedge j_1+\kappa\,j_3\wedge j_3
-2\chi\,j_2\wedge j_2-\chi\,(j_1\wedge j_3+j_3\wedge j_1)&=0\,,\\[2mm]
(\kappa\,j_1+*j_1)\wedge j_2+j_2\wedge(\kappa\,j_1+*j_1)
+\chi\,(j_2\wedge j_3+j_3\wedge j_2)&=0\,,\\[2mm]
(\kappa\,j_3-*j_3)\wedge j_2+j_2\wedge(\kappa\,j_3-*j_3)
-\chi\,(j_2\wedge j_1+j_1\wedge j_2)&=0\,.
\end{aligned}
\end{equation}
The terms proportional to $\chi$ arise from the Wess--Zumino term, whose variation is indeed local. Together with the Maurer--Cartan equations~\eqref{eq:MC-Z4}, the equations of motion determine the classical dynamics of the supercoset sigma model. In conformal gauge they are supplemented by the Virasoro constraints, which we shall not need explicitly. Since $\sqrt\lambda$ multiplies the full action, the classical equations depend on the background only through the parameter $\chi$.

\subsection{Classical integrability of the supercoset sigma model}\label{sec:lax}

Classical integrability of string theory in AdS backgrounds was first established for the GS superstring on AdS$_5\times$S$^5$. Bena, Polchinski and Roiban constructed a one-parameter family of flat connections for the supercoset sigma model, whose flatness for all values of the spectral parameter is equivalent to the equations of motion \cite{Bena:2003wd}. The construction only relies on the $\mathbb{Z}_4$ grading of the superalgebra and therefore applies to all semi-symmetric superspaces \cite{Adam:2007ws,Zarembo:2010sg}. For AdS$_3\times$S$^3$ with pure RR flux, the corresponding flat connection was found in \cite{Chen:2005uj,Babichenko:2009dk}. 
The extension to mixed flux is due to Cagnazzo and Zarembo, who constructed the Wess--Zumino term~\eqref{eq:H} and showed that the Lax connection survives its inclusion, provided that the two couplings obey the relation in equation~\eqref{eq:chi-kappa} \cite{Cagnazzo:2012se}. This generalises the classical integrability of the bosonic principal chiral model with a Wess--Zumino term \cite{Veselov:1984}.

Following reference~\cite{Cagnazzo:2012se}, we consider the ansatz
\begin{equation}\label{eq:lax}
\Lax(x)=j_0+a(x)\,j_2+b(x)\,{*j_2}+c_1(x)\,j_1+c_3(x)\,j_3
\end{equation}
for a connection that depends on a spectral parameter $x$. Since $j_0$ transforms as a connection and the remaining currents transform in the adjoint representation, see equation~\eqref{eq:gauge-Z4}, the connection $\Lax(x)$ transforms as a gauge field under the coset symmetry. Inserting the ansatz into the flatness condition
\begin{equation}\label{eq:flatness}
d\Lax(x)+\Lax(x)\wedge\Lax(x)=0
\end{equation}
and using the Maurer--Cartan equations~\eqref{eq:MC-Z4} and the equations of motion~\eqref{eq:eom}, one obtains an overdetermined system of algebraic equations for the four coefficients. Cagnazzo and Zarembo showed that this system admits non-trivial solutions only if the couplings satisfy the relation in equation~\eqref{eq:chi-kappa} \cite{Cagnazzo:2012se}. Integrability thus singles out the same relation between $\chi$ and $\kappa$ as the supergravity equations, kappa symmetry and conformal invariance. Once the relation holds, the system is equivalent to the four conditions
\begin{equation}\label{eq:coefficient-identities}
c_1^2=a-\kappa\,b\,,\qquad c_3^2=a+\kappa\,b\,,\qquad
c_1c_3=1-\chi\,b\,,\qquad a^2-b^2+2\chi\,b=1\,,
\end{equation}
where the last relation follows from the first three by means of equation~\eqref{eq:chi-kappa}. The conditions therefore leave one of the coefficients undetermined. It plays the role of the spectral parameter. Conversely, if the flatness condition holds for all values of the spectral parameter, it implies both the Maurer--Cartan equations and the equations of motion. Hence the supercoset sigma model with mixed flux is classically integrable.

A convenient choice of spectral parameter is one for which $a$, $b$ and the squares of $c_1$ and $c_3$ are rational functions. Following reference~\cite{Cagnazzo:2012se}, we use
\begin{equation}\label{eq:lax-coeff}
\begin{aligned}
a(x)&=\kappa\,\frac{x^2+1}{x^2-1}\,,&\qquad
b(x)&=\chi-\frac{2\kappa\,x}{x^2-1}\,,\\[2mm]
c_1(x)&=\frac{\sqrt{\kappa(1-\chi)}\,x+\sqrt{\kappa(1+\chi)}}{\sqrt{x^2-1}}\,,
&\qquad
c_3(x)&=\frac{\sqrt{\kappa(1+\chi)}\,x-\sqrt{\kappa(1-\chi)}}{\sqrt{x^2-1}}\,,
\end{aligned}
\end{equation}
where a branch of the square root $\sqrt{x^2-1}$ has been fixed. One may verify directly that these coefficients satisfy the conditions in equation~\eqref{eq:coefficient-identities}. In components along $d\tau$ and $d\sigma$, the Lax connection reads
\begin{equation}\label{eq:lax-components}
\begin{aligned}
\Lax^\tau(x)&=j_0^\tau+a(x)\,j_2^\tau+b(x)\,j_2^\sigma
+c_1(x)\,j_1^\tau+c_3(x)\,j_3^\tau\,,\\[2mm]
\Lax^\sigma(x)&=j_0^\sigma+a(x)\,j_2^\sigma+b(x)\,j_2^\tau +c_1(x)\,j_1^\sigma+c_3(x)\,j_3^\sigma\,,
\end{aligned}
\end{equation}
where we used the conventions in equation~\eqref{eq:orientation} for the Hodge star. For pure RR flux, i.e.\ at $\chi=0$ and $\kappa=1$, the coefficients reduce to
\begin{equation}\label{eq:lax-RR}
a^\mathrm{RR}(x)=\frac{x^2+1}{x^2-1}\,,\qquad b^\mathrm{RR}(x)=-\frac{2x}{x^2-1}\,,\qquad
c^\mathrm{RR}_1(x)=\sqrt{\frac{x+1}{x-1}}\,,\qquad c^\mathrm{RR}_3(x)=\sqrt{\frac{x-1}{x+1}}\,,
\end{equation}
which is the Lax connection of Bena, Polchinski and Roiban adapted to the supercoset $\mathcal{M}$ \cite{Bena:2003wd,Babichenko:2009dk}. In the opposite limit of pure NSNS flux, the coefficients $a$, $c_1$ and $c_3$ tend to zero while $b$ tends to one. The Lax connection then becomes independent of the spectral parameter and degenerates. This limit corresponds to the WZW point, where the chiral factorisation of the currents takes over the role of integrability \cite{Cagnazzo:2012se}.

\smallskip

The goal now is to extract conserved quantities from the Lax connection. To this
end, one introduces the usual transport matrix
\begin{equation}\label{eq:transport}
T(\sigma_2,\sigma_1;x)=\mathcal{P}\exp\Big[-\int_{\sigma_1}^{\sigma_2} d\sigma\,\Lax^\sigma(\tau,\sigma;x)\Big]\,,
\end{equation}
where $\mathcal{P}$ denotes path ordering. As a consequence of the flatness condition~\eqref{eq:flatness}, its time evolution only depends on the Lax connection at the endpoints,
\begin{equation}\label{eq:transport-evolution}
\partial_\tau T(\sigma_2,\sigma_1;x) =-\Lax^\tau(\sigma_2;x)\,T(\sigma_2,\sigma_1;x)
+T(\sigma_2,\sigma_1;x)\,\Lax^\tau(\sigma_1;x)\,.
\end{equation}
For a closed string, the fields, and hence also the Lax connection itself, are periodic with period $2\pi$, so that the monodromy matrix $T(2\pi,0;x)$ evolves by conjugation. Hence, its eigenvalues are conserved for all values of the spectral parameter, and their expansion in $x$ provides an infinite set of conserved charges. For an open string on the strip, the two Lax matrices that appear in the first and second terms of equation~\eqref{eq:transport-evolution} need not be the same. Conserved quantities exist only if the boundary conditions at $\sigma=0$ and $\sigma=\pi$ can be lifted to gluing conditions for the Lax connection, in which case the monodromy matrix is replaced by a double-row monodromy matrix \cite{Sklyanin:1988yz,Dekel:2011ja}. The boundary conditions for AdS$_2$ branes are derived in section~\ref{sec:branes}, and their lift to the Lax connection is constructed in section~\ref{sec:integrability}.

\section{\texorpdfstring{$\textrm{AdS}_2$}{AdS2} branes and boundary conditions}\label{sec:branes}

In this section we introduce the branes whose integrability we study, and we derive the boundary conditions they impose on the supercoset sigma model. We first describe the branes from the spacetime perspective and determine their worldvolume fields for all values of the fluxes. We then construct the boundary action of the open string and derive the boundary conditions from its variation. At the end of the section we show that the resulting gluing conditions reproduce the worldvolume fields.

\subsection{\texorpdfstring{$\textrm{AdS}_2 \times \textrm{S}^2$}{AdS2 x S2} branes -- the spacetime perspective}\label{sec:spacetime-branes}

The branes we consider are D3-branes that wrap an AdS$_2$ subspace of AdS$_3$ and an S$^2$ subspace of S$^3$, and that sit at a point of the torus. Open strings ending on them satisfy (generalized) Neumann conditions along AdS$_2\times$S$^2$, and Dirichlet conditions along the two transverse directions in AdS$_3\times$S$^3$ and along T$^4$. For pure NSNS flux, these branes are well understood. In the WZW model on $SL(2,\mathbb{R})\times SU(2)$, they are products of spherical S$^2$ branes that are localised along the conjugacy classes of $SU(2)\cong$ S$^3$ \cite{Alekseev:1998mc} and AdS$_2$ branes that are localised along twisted conjugacy classes of $SL(2,\mathbb{R}) \cong$ AdS$_3$ \cite{Bachas:2000fr}. Our aim now is to describe these branes for the whole range of fluxes of the bulk theory.

To get there let us be a bit more specific about the embedding of the branes into the bosonic target of the sigma model. Using the group manifold description of AdS$_3\times$S$^3$ from subsection~\ref{sec:background}, we realise $SU(1,1)\cong SL(2,\mathbb{R})$ by real matrices. We parametrise the two blocks of a bosonic group element $g=\mathrm{diag}(g_A,g_S)$ as
\begin{equation}\label{eq:brane-coordinates}
\begin{aligned}
g_A(\psi,\rho,t)&=\sinh\psi\,\sigma_3+\cosh\psi\,e^{it\sigma_2/2}e^{\rho\sigma_1}
e^{it\sigma_2/2}\,,\\[2mm]
g_S(\alpha,\beta,\gamma)&=e^{-i\gamma\sigma_3/2}e^{i\beta\sigma_1/2}
 e^{i\alpha\sigma_3} e^{-i\beta\sigma_1/2}e^{i\gamma\sigma_3/2}\,.
\end{aligned}
\end{equation}
In these coordinates, the metric of unit-radius AdS$_3\times$S$^3$ takes the form
\begin{equation}\label{eq:slicing}
\begin{aligned}
ds^2_{\AdS_3}&=d\psi^2+\cosh^2\psi\,\big(-\cosh^2\!\rho\,dt^2+d\rho^2\big)\,,\\[2mm]
ds^2_{\Sph^3}&=d\alpha^2+\sin^2\!\alpha\,\big(d\beta^2+\sin^2\!\beta\,d\gamma^2\big)\,.
\end{aligned}
\end{equation}
In the background of radius $R$, hypersurfaces of constant $\psi\in\mathbb{R}$ are copies of AdS$_2$ with radius $R\cosh\psi$, and surfaces of constant $\alpha\in(0,\pi)$ are two-spheres with radius $R\sin\alpha$. It is easy to see that these transverse positions can be characterised by the traces
\begin{equation}\label{eq:brane-traces}
\tr(w_A\,g_A)=2i\sinh\psi\,,\qquad \tr(g_S)=2\cos\alpha\,,\qquad w=\mathrm{diag}(w_A,w_S)=\mathrm{diag}(i\sigma_3,\mathbf{1}_2)\,.
\end{equation}
Hence the S$^2$ is a conjugacy class of $SU(2)$, while the AdS$_2$ is a twisted conjugacy class of $SL(2,\mathbb{R})$ as we had mentioned before. The twist is the conjugation with $\sigma_3$, which is an outer automorphism of $SL(2,\mathbb{R})$ since $\sigma_3$ has determinant $-1$ \cite{Bachas:2000fr}. The conditions that arise from setting the traces in equation~\eqref{eq:brane-traces} to some constant values are invariant under the twisted conjugations
\begin{equation}\label{eq:twisted-conjugation}
w\,g\longmapsto h\,(w\,g)\,h^{-1}\,,\qquad h\in SL(2,\mathbb{R})\times SU(2)\,.
\end{equation}
Hence, these transformations are not only isometries of AdS$_3\times$S$^3$ but they also restrict to isometries of the worldvolume AdS$_2\times$S$^2$. In the supersymmetric theory, the twisted conjugation extends to the supergroup. The brane then preserves a subgroup $PSU(1,1|2)$ of the superisometry group in equation~\eqref{eq:superisometry}, which is embedded through the twisted diagonal. Hence the brane preserves eight of the sixteen supercharges of the background, i.e.\ it is half-BPS. We shall see in subsection~\ref{sec:boundary-conditions} that the boundary conditions respect this symmetry for all values of the fluxes.
\medskip

The brane carries a worldvolume gauge field, whose gauge-invariant field strength
\begin{equation}\label{eq:worldvolume-field}
\mathcal{F}=B+2\pi\alpha'f\,,\qquad
\mathsf{e}=\mathcal{F}_{01}\,,\qquad \mathsf{b}=\mathcal{F}_{23}\,,
\end{equation}
combines the pullback of the NSNS two-form $B$ with the field strength $f$ of the gauge field.\footnote{Let us note that the split of $\mathcal{F}$ into $B$ and $2\pi \alpha' f$ is not unique since the local two-form potential $B$ of the NSNS three-form $\text{H}$ is not unique. Later we shall make a specific choice of $B$.} Since $\mathcal{F}$ must be invariant under the isometries of AdS$_2\times$S$^2$, and since both factors are two-dimensional, the field strength $\mathcal{F}$ has only two (constant) components in orthonormal frames $(0,1)$ on AdS$_2$ and $(2,3)$ on S$^2$. We shall denote the electric field on AdS$_2$ by $\mathsf{e}$ and the magnetic field on S$^2$ by $\mathsf{b}$. For pure NSNS flux, the fields take the values \cite{Alekseev:1998mc,Bachas:2000fr,Bachas:2000ik}
\begin{equation}\label{eq:fields-NSNS}
\mathsf{e}_0=\tanh\psi\,,\qquad \mathsf{b}_0=\cot\alpha\,,
\end{equation}
with the orientations specified in appendix~\ref{app:brane-eom}. These fields are needed to stabilise the brane at generic positions \cite{Bachas:2000ik,Bachas:2000fr}. At the quantum level, the magnetic flux through the S$^2$ is quantised, which restricts $\alpha$ to discrete values \cite{Alekseev:1998mc}. In our classical analysis, both positions are treated as continuous parameters. For $\psi=0$ and $\alpha=\pi/2$ the fields vanish, and the brane wraps an AdS$_2$ through the centre of AdS$_3$ together with a maximal S$^2$. In the remainder of this subsection we show that branes with the same embedding exist for all values of the fluxes, and we determine their worldvolume fields.

The dynamics of the brane is governed by the action
\begin{equation}\label{eq:D3-action}
S_{\mathrm{D3}}=-T_3\int d^4\xi\,e^{-\Phi}\sqrt{-\det\big(G+\mathcal{F}\big)}-T_3\int\big(C_4+C_2\wedge\mathcal{F}\big)\,,
\end{equation}
where $G$ is the induced metric and $T_3e^{-\Phi}$ is the tension of the D3-brane. The first term is the Dirac--Born--Infeld action, and the second term is the Wess--Zumino coupling to the RR potentials, in which we used that the RR axion vanishes. Since the induced metric is diagonal and $\mathcal{F}$ is block-diagonal, the Dirac--Born--Infeld Lagrangian per unit induced volume, in units of $T_3e^{-\Phi}$, reads
\begin{equation}\label{eq:DBI}
L(\mathsf{e},\mathsf{b}) =-\sqrt{(1-\mathsf{e}^2)(1+\mathsf{b}^2)}\,.
\end{equation}
As in nonlinear electrodynamics, we define the dual fields
\begin{equation}\label{eq:dual-fields}
\mathsf{d}=\frac{\partial L}{\partial\mathsf{e}}
=\mathsf{e}\,\frac{\sqrt{1+\mathsf{b}^2}}{\sqrt{1-\mathsf{e}^2}}\,,\qquad
\mathsf{h}=-\frac{\partial L}{\partial\mathsf{b}}
=\mathsf{b}\,\frac{\sqrt{1-\mathsf{e}^2}}{\sqrt{1+\mathsf{b}^2}}\,.
\end{equation}
For the pure NSNS brane, the fields in equation~\eqref{eq:fields-NSNS} give the dual fields
\begin{equation}\label{eq:dual-fields-NSNS}
\mathsf{d}_0=\frac{\sinh\psi}{\sin\alpha}\,,\qquad \mathsf{h}_0=\frac{\cos\alpha}{\cosh\psi}\,.
\end{equation}
Constant fields solve the equations of motion of the gauge field, since the pullbacks of $\text{H}$ and $F_3$ to the brane vanish.

The positions $\psi$ and $\alpha$ are determined by the equations of motion of the embedding. To derive them, we vary the position of the brane at fixed gauge field. Such a variation changes the induced metric, and through the pullback of $B$ it shifts the field strength $\mathcal{F}$ by the contraction of $\text{H}$ with the variation. The variation of the Wess--Zumino term can be expressed through the gauge-invariant RR field strengths. Since the five-form flux vanishes, it reduces to the contraction of $F_3$ with the variation, wedged with $\mathcal{F}$, up to terms that vanish by the equations of motion of the gauge field. Evaluating these contributions in the background of equation~\eqref{eq:nh-background}, we obtain the equations of motion for $\psi$ and $\alpha$,
\begin{equation}\label{eq:force-balance}
\varepsilon\tanh\psi=\chi\,\mathsf{d}-\kappa\,\mathsf{b}\,,\qquad
\varepsilon^{-1}\cot\alpha=\chi\,\mathsf{h}+\kappa\,\mathsf{e}\,,\qquad
\varepsilon=\mathsf{e}\,\mathsf{d}-L=\frac{\sqrt{1+\mathsf{b}^2}}{\sqrt{1-\mathsf{e}^2}}\,,
\end{equation}
where $\varepsilon$ is the energy density of the brane per unit induced volume. Further details on the derivation of the equations of motion are provided in appendix~\ref{app:brane-eom}. The left-hand sides of equation~\eqref{eq:force-balance} arise from the variation of the induced metric, the terms proportional to $\chi$ from the shift of $\mathcal{F}$, and the terms proportional to $\kappa$ from the Wess--Zumino term. For the opposite orientation of the brane, the sign of $\kappa$ is reversed. The first equation balances the tension of the brane, which pulls it towards $\psi=0$, against the forces that the NSNS and RR fluxes exert on the dissolved fundamental strings and D1-branes, whose densities are $\mathsf{d}$ and $\mathsf{b}$. With the relations in equation~\eqref{eq:dual-fields}, the two equations take the simple form
\begin{equation}\label{eq:brane-eom}
\chi\,\mathsf{e}-\kappa\,\mathsf{h}=\tanh\psi\,,\qquad
\chi\,\mathsf{b}+\kappa\,\mathsf{d}=\cot\alpha\,.
\end{equation}
For pure NSNS flux, they reproduce the fields in equation~\eqref{eq:fields-NSNS}. As they stand, these equations relate the fields $\mathsf{e}$ and $\mathsf{b}$ to the position parameters $\psi$ and $\alpha$ of the brane, with coefficients given by the background fluxes $\chi$ and $\kappa$. While they look linear, one should keep in mind that the dual fields $\mathsf{d}$ and $\mathsf{h}$ are related to $\mathsf{e}$ and $\mathsf{b}$ through equation~\eqref{eq:dual-fields}.

Exploiting the linearity of the equations~\eqref{eq:brane-eom} in the pairs $(\mathsf{e},\mathsf{h})$ and $(\mathsf{b},\mathsf{d})$, they can be solved with the help of the duality rotations of Gaillard and Zumino \cite{Gaillard:1981rj}. For our fields, a duality rotation by an angle $\vartheta$ acts as
\begin{equation}\label{eq:GZ-rotation}
\begin{aligned}
\mathsf{e}'&=\cos\vartheta\,\mathsf{e}+\sin\vartheta\,\mathsf{h}\,,&\qquad
\mathsf{h}'&=\cos\vartheta\,\mathsf{h}-\sin\vartheta\,\mathsf{e}\,,\\[2mm]
\mathsf{b}'&=\cos\vartheta\,\mathsf{b}-\sin\vartheta\,\mathsf{d}\,,&\qquad
\mathsf{d}'&=\cos\vartheta\,\mathsf{d}+\sin\vartheta\,\mathsf{b}\,.
\end{aligned}
\end{equation}
The rotated fields obey the same constitutive relations as the original ones, i.e.\ they are related as in equation~\eqref{eq:dual-fields}, provided that the condition
\begin{equation}\label{eq:GZ-condition}
\mathsf{e}\,\mathsf{b}=\mathsf{d}\,\mathsf{h}
\end{equation}
holds for all values of the fields. This is the criterion of Gaillard and Zumino \cite{Gaillard:1981rj}, specialised to our configuration. It is satisfied by the fields in equation~\eqref{eq:dual-fields}, which reflects the duality covariance of the Born--Infeld action \cite{Gibbons:1995cv}. We now choose the angle
\begin{equation}\label{eq:rotation-angle}
\cos\vartheta=\chi\,,\qquad \sin\vartheta=\kappa\,.
\end{equation}
According to equation~\eqref{eq:brane-eom}, the rotation by $-\vartheta$ maps the fields of the brane to those of the pure NSNS brane in equation~\eqref{eq:fields-NSNS}. Hence it maps the dual fields to those in equation~\eqref{eq:dual-fields-NSNS}, and the worldvolume fields follow by rotating the fields of the pure NSNS brane by $\vartheta$,
\begin{equation}\label{eq:rotated-fields}
\mathsf{e}=\chi\,\mathsf{e}_0+\kappa\,\mathsf{h}_0
=\chi\tanh\psi+\kappa\,\frac{\cos\alpha}{\cosh\psi}\,,\qquad
\mathsf{b}=\chi\,\mathsf{b}_0-\kappa\,\mathsf{d}_0
=\chi\cot\alpha-\kappa\,\frac{\sinh\psi}{\sin\alpha}\,.
\end{equation}
These are the worldvolume fields of the brane for all values of the fluxes and all positions. By the Cauchy--Schwarz inequality, the magnitude of the electric field stays below one, as required for the Dirac--Born--Infeld action to be real.

The duality rotation has a simple origin. For the D3-brane, it is the worldvolume realisation of the $SL(2,\mathbb{R})$ symmetry of type IIB supergravity \cite{Tseytlin:1996it,Green:1996qg}. It is accompanied by the same rotation of the background doublet $(\text{H},g_sF_3)$, while the Einstein-frame metric, and with it the embedding of the brane, is left invariant. As noted in subsection~\ref{sec:background}, the mixed flux background in equation~\eqref{eq:nh-background} is obtained from the pure NSNS one by the rotation with the angle in equation~\eqref{eq:rotation-angle}. The fields in equation~\eqref{eq:rotated-fields} show that the brane is obtained from the pure NSNS brane by the same rotation. In particular, at pure RR flux the brane carries the dual fields of the pure NSNS brane, with a sign for the magnetic field, in agreement with S-duality. At $\psi=0$ and $\alpha=\pi/2$ all four fields vanish, and the brane is unaffected by the rotation. The simple form of the solution relies on the self-duality of the D3-brane, which singles out the brane that wraps AdS$_2\times$S$^2$ at a point of T$^4$.

\subsection{Boundary conditions in the supercoset sigma model}\label{sec:boundary-conditions}

We now derive the boundary conditions from the supercoset sigma model of subsection~\ref{sec:coset}. In the case of an open worldsheet $M$, the definition of the Wess--Zumino term requires specifying additional geometric data. We refer to the original papers~\cite{Alekseev:1998mc,Gawedzki:2001ye} for a detailed discussion. Since the boundary conditions are local, we  may assume without loss of generality that the boundary $\partial M$ of $M$ has a single connected component. Following \cite{Gawedzki:2001ye}, we extend the sigma model fields in two steps. First we glue in a disc $\mathcal{D}$ into $M$ in order to construct a closed surface $M'$.\footnote{We oriented the disc such that its boundary agrees with $\partial M$, and we glue in $\mathcal{D}$ with opposite orientation.} The sigma model field is extended to the disc such that it takes values in a conjugacy class. Then we choose a three-manifold $\mathcal{B}$ whose boundary is $M'$ and extend the sigma model field to $\mathcal{B}$. In formulas, we have
\begin{equation}\label{eq:cap}
\partial\mathcal{D}=\partial M\,,\qquad M'=M-\mathcal{D}\,,\qquad
\partial\mathcal{B}=M'\,.
\end{equation}
There is a lot of freedom in these two extensions, but the action
\begin{equation}\label{eq:open-action}
S=\frac{\sqrt\lambda}{2\pi}\int_M\str\big(j_2\wedge*j_2+\kappa\,j_1\wedge j_3\big)+\frac{\sqrt\lambda}{2\pi}\Big(\chi\int_{\mathcal{B}}H+\int_{\mathcal{D}}\mathcal{F}\Big)\,,
\end{equation}
may easily be seen to be invariant under continuous deformations of the choices we made, provided that the two-form $\mathcal{F}$ on the brane satisfies
\begin{equation}\label{eq:dF}
d\mathcal{F}=\chi\,\iota^*H\,.
\end{equation}
Invariance under large deformations leads to the usual quantisation conditions \cite{Alekseev:1998mc,Gawedzki:2001ye}. For bosonic configurations, the two-form $\mathcal{F}$ is the worldvolume field strength in equation~\eqref{eq:worldvolume-field}, measured in units of $R^2$.
\smallskip

In order to construct the two-form $\mathcal{F}$, we first determine a two-form potential $\omega$ for $H$ on the brane. To this end, we rewrite the three-form $H$ by expressing it through the gauge-invariant combination
\begin{equation}\label{eq:g}
g=g_L\,g_R^{-1}\,,\qquad g^{-1}dg=\Ad_{g_R}\big(j_L-j_R\big)\,,\qquad \Ad_gX=gXg^{-1}\,,
\end{equation}
where $\Ad_g$ denotes the usual adjoint action. For bosonic configurations, $g$ is the point of AdS$_3\times$S$^3\cong SU(1,1)\times SU(2)$ that we parametrised in equation~\eqref{eq:brane-coordinates}, and in the gauge $g_R=1$ it reduces to the field in equation~\eqref{eq:bosonic-gauge}. As we show in appendix~\ref{app:potential}, the three-form in equation~\eqref{eq:H} can be written as
\begin{equation}\label{eq:H-g}
H=\frac16\,\str\big(g^{-1}dg\wedge g^{-1}dg\wedge g^{-1}dg\big) -i\,d\,\str\big(j_1\wedge j_3\big)\,.
\end{equation}
Thus $H$ is the Wess--Zumino three-form of the gauge-invariant field $g$, up to an exact term that is built from the fermionic currents.

The brane of subsection~\ref{sec:spacetime-branes} extends to the supergroup as the set of elements $g$ for which $w\,g$ lies in a fixed conjugacy class,
\begin{equation}\label{eq:brane}
w\,g=s\,u\,s^{-1}\,,\qquad s\in G_R\,.
\end{equation}
Here $u=\mathrm{diag}(u_A,u_S)$ is a fixed bosonic element of $w\,G^B$, whose traces are related to the position of the brane through
\begin{equation}\label{eq:u-traces}
\tr(u_A)=2i\sinh\psi\,,\qquad \tr(u_S)=2\cos\alpha\,,
\end{equation}
where $\psi$ and $\alpha$ are the coordinates introduced in equation~\eqref{eq:brane-coordinates}. 
Conjugation with $w$ preserves the supertrace and maps $G_L$ to $G_R$, since $w^\dagger H_Rw=H_L$, see equation~\eqref{eq:real}.

On the tangent directions to the conjugacy class, the three-form $H$ possesses the two-form potential
\begin{equation}\label{eq:omega}
\omega=\frac14\,\str\Big(g^{-1}dg\wedge\frac{1+\Ad_{wg}}{1-\Ad_{wg}}\,g^{-1}dg\Big)-i\,\str\big(j_1\wedge j_3\big)\,,\qquad d\omega=\iota^*H\,,
\end{equation}
where all forms are restricted to the brane. The first term generalises the two-form that was found in reference~\cite{Alekseev:1998mc} for conjugacy classes in WZW models, while the second term is needed to reproduce the exact term in equation~\eqref{eq:H-g}. The proof of the last relation in equation~\eqref{eq:omega} is given in appendix~\ref{app:potential}.
\smallskip

Now that we have determined one two-form potential $\omega$ of the field strength $H$ on the brane we can write the most general solution of the condition in equation~\eqref{eq:dF} as
\begin{equation}\label{eq:F-general}
\mathcal{F}=\chi\,\omega+\mathcal{F}_0\,,\qquad d\mathcal{F}_0=0\,.
\end{equation}
The closed two-form $\mathcal{F}_0$ is the remaining freedom in the boundary action, and its choice will be crucial below. Before we fix $\mathcal{F}_0$, we want to determine the boundary conditions that the fields must satisfy in order for the action~\eqref{eq:open-action} to be well defined.

To this end, we vary the action as in our discussion of the bulk theory, see equation~\eqref{eq:variations}. Since $H$ is closed and $\mathcal{F}$ obeys the condition in equation~\eqref{eq:dF}, Cartan's formula
\begin{equation}\label{eq:cartan}
\delta X=\iota_\Delta(dX)+d\bigl(\iota_\Delta X\bigr)
\end{equation}
for the variation of a form $X$ gives
\begin{equation}\label{eq:cap-variation}
\delta\Big(\chi\int_{\mathcal{B}}H+\int_{\mathcal{D}}\mathcal{F}\Big)
=\chi\int_M\iota_\Delta H+\int_{\partial M}\iota_\Delta\mathcal{F}\,,
\end{equation}
where $\iota_\Delta$ denotes the contraction with the variation. The first term contributes to the bulk equations of motion~\eqref{eq:eom}, while the second term only depends on the fields at the boundary. Together with the boundary terms that arise from the kinetic and GS terms, originally computed in \cite{Dekel:2011ja}, equation~\eqref{eq:cap-variation} gives rise to the boundary variation
\begin{equation}\label{eq:boundary-variation}
\begin{aligned}
\delta S\big|_{\partial M}=-\frac{\sqrt\lambda}{2\pi}\int d\tau\,\Big[\,
&2\str\big(\Delta_2\,j_2^\sigma\big)
+(\kappa-i\chi)\str\big(\Delta_1\,j_3^\tau-\Delta_3\,j_1^\tau\big)\\[2mm]
& \hspace{-5mm} +\frac{\chi}{2}\str\Big((\Delta_L-\Delta_R)\,\frac{1+\Ad_{\hat u}}{1-\Ad_{\hat u}}\, \big(j_L^\tau-j_R^\tau\big)\Big) +\mathcal{F}_0(\Delta,\partial_\tau)\Big]_{\sigma=0}^{\sigma=\pi}\,,
\end{aligned}
\end{equation}
where $\mathcal{F}_0(\Delta,\partial_\tau)$ is the contraction of $\mathcal{F}_0$ with the boundary variation and we introduced
\begin{equation}\label{eq:u-hat}
\hat u=g_R^{-1}\,w\,g_L\,.
\end{equation}
The boundary variation is gauge invariant, since $\hat u$ transforms by conjugation under gauge transformations. For configurations on the superbrane, see equation~\eqref{eq:superbrane} below, the trace relations in
 equation~\eqref{eq:u-traces} therefore also apply to $\hat u$.

Before we analyse the required boundary conditions further in the full theory it is actually quite instructive to consider purely bosonic configurations, with vanishing $j_1$ and $j_3$. For the moment we shall also set $\mathcal{F}_0$ to zero. In the gauge $g_R=1$, the currents, the variations and the element $\hat u$ then become
\begin{equation}\label{eq:bosonic-gauge-boundary}
j_L=2j_2\,,\qquad \Delta_L=2\Delta_2\,,\qquad \hat u=w\,g\,
\end{equation}
so that the boundary variation~\eqref{eq:boundary-variation} vanishes provided that
\begin{equation}\label{eq:bosonic-bc}
j_2^\sigma=-\chi\,\frac{1+\Ad_{wg}}{1-\Ad_{wg}}\,j_2^\tau
\end{equation}
along the directions tangent to the brane, while $j_2^\tau=0$ in the two normal directions, which commute with $w\,g$. In terms of the light-cone components
\begin{equation}\label{eq:light-cone}
j_{2,\pm}=j_2^\tau\pm j_2^\sigma\,,
\end{equation}
the tangential condition reads
\begin{equation}\label{eq:bosonic-bc-lc}
\big[(1+\chi)-(1-\chi)\Ad_{wg}\big]\,j_{2,+}
=\big[(1-\chi)-(1+\chi)\Ad_{wg}\big]\,j_{2,-}\,.
\end{equation}
For $\chi=1$, this reduces to the gluing condition
\begin{equation}\label{eq:WZW-gluing}
j_{2,+}=-\Ad_{wg}\,j_{2,-}
\end{equation}
for twisted conjugacy classes in the WZW model \cite{Alekseev:1998mc,Bachas:2000fr}, while for $\chi=0$ it imposes the usual Neumann conditions along the brane.

In the purely bosonic theory there is, however, considerable freedom. Indeed, the three-form $H$ has separate contributions from AdS$_3$ and S$^3$, and each pulls back to zero on the corresponding two-dimensional factor of the AdS$_2\times$S$^2$ brane. Therefore, the pullback $\iota^*H$ vanishes and $\omega$ is closed, see equation~\eqref{eq:omega}. For this reason, we may also choose $\mathcal{F}_0=(1-\chi)\,\omega$, which turns the boundary condition~\eqref{eq:bosonic-bc} into the usual WZW gluing condition for any value of the parameter $\chi$. This is the choice that was adopted in~\cite{Gil:2026ost} and it was shown there that it gives rise to integrable boundary conditions. Once fermions are included, however, we can no longer work with this $\mathcal{F}_0$, since $\iota^*H$ no longer vanishes and hence $\omega$ is not closed on the superbrane. As we shall see, a different choice of $\mathcal{F}_0$ is allowed and will play a crucial role in the fermionic theory.
\smallskip

So, let us now return to the general case with nonvanishing fermions. In the supersymmetric theory, we require that the boundary values of the fields satisfy
\begin{equation}\label{eq:superbrane}
\hat u=h\,u\,h^{-1}\,,\qquad h\in G^B\,, \qquad \partial_\tau u=0\,,
\end{equation}
i.e.\ the element $\hat u$ in equation~\eqref{eq:u-hat} must be conjugate to a constant $u$ by an element of the bosonic subgroup. This implies the condition~\eqref{eq:brane}, with $s=g_Rh$. It is invariant under gauge transformations and under the twisted diagonal action of $PSU(1,1|2)$. Using a gauge transformation, we can choose the representative
\begin{equation}\label{eq:representative}
(g_L,g_R)=(w^{-1}s\,u\,,\,s)\,,
\end{equation}
which satisfies the condition in equation~\eqref{eq:brane}, with constant $\hat u=u$. Along the brane, the
currents then read
\begin{equation}\label{eq:brane-currents}
j_R=\theta\,,\qquad j_L=\Ad_u^{-1}\theta\,,\qquad \theta=s^{-1}ds\,.
\end{equation}
For the odd components, these relations give the fermionic boundary conditions,
\begin{equation}\label{eq:fermionic-bc}
j_R^{F,\tau}=\Ad_{\hat u}\,j_L^{F,\tau}\,,\qquad
\Delta_R^F=\Ad_{\hat u}\,\Delta_L^F\,.
\end{equation}
Here we have reinstated $\hat u$ to write them in a manifestly gauge-covariant form.
In terms of the $\mathbb{Z}_4$ components, the first condition becomes
\begin{equation}\label{eq:fermionic-bc-Z4}
\big(1-i\Ad_{\hat u}\big)\,j_1^\tau=\big(1+i\Ad_{\hat u}\big)\,j_3^\tau\,.
\end{equation}
An analogous condition holds for the variations. The fermionic boundary conditions do not depend on the fluxes. They simply express that the fermions at the boundary take values in the superbrane. For the even components, the restriction to the brane gives Dirichlet conditions along the two normal directions, while the tangent components obey
\begin{equation}\label{eq:even-tangent}
2\Delta_2=\big(\Ad_u^{-1}-1\big)\Delta_R^B\,,\qquad 2j_2^\tau=\big(\Ad_u^{-1}-1\big)j_R^{B,\tau}\,.
\end{equation}
These relations follow from equation~\eqref{eq:brane-currents} and the definition of $j_2$. In the representative in equation~\eqref{eq:representative}, the fermionic contributions to $\omega$ cancel, see
appendix~\ref{app:potential}, and the potential reduces to
\begin{equation}\label{eq:omega-brane}
\omega=\frac12\,\str\big(\theta^B\wedge\Ad_u\theta^B\big)\,.
\end{equation}
Hence the terms in the boundary variation~\eqref{eq:boundary-variation} that are proportional to $\chi$ contain no fermions. The GS term, however, leaves a fermionic contribution. Using equation~\eqref{eq:brane-currents}, one finds
\begin{equation}\label{eq:odd-residual}
\kappa\,\str\big(\Delta_1\,j_3^\tau-\Delta_3\,j_1^\tau\big)
=-\frac{i\kappa}{2}\,\str\Big(\Delta_R^F\,\big(\Ad_u-\Ad_u^{-1}\big)\,j_R^{F,\tau}\Big)\,.
\end{equation}
This term vanishes for pure NSNS flux, and also for special classes on which $\Ad_u$ squares to the identity on the odd subspace. In general, however, it is nonzero. Since the fermionic boundary conditions~\eqref{eq:fermionic-bc} already relate the left and right fermions, the remaining variations $\Delta_R^F$ are arbitrary. The contribution in equation~\eqref{eq:odd-residual} must therefore be cancelled by a suitable choice of $\mathcal{F}_0$.

Having worked out some of the equations that constrain $\mathcal{F}_0$ we now want to spell out the choice we propose to adopt:
\begin{equation}\label{eq:F0a}
\mathcal{F}_0=\str\big(t_0\,\theta\wedge\theta\big)=-d\,\str(t_0\,\theta)\,,
\end{equation}
with the matrix $t_0=\mathrm{diag}\big(t_{0,A},t_{0,S}\big)$ built from the blocks
\begin{equation} \label{eq:F0b}
t_{0,I}=-\frac{i\kappa}{4}\,\tr(u_J)\,\big(u_I-u_I^{-1}\big)\,,
\end{equation}
where $(I,J)=(A,S)$ or $(S,A)$. Since $t_0$ is constant and commutes with the stabiliser of $u$, $\mathcal{F}_0$ descends to a closed two-form on the brane. Note that $t_{0,I}$ involves a trace in the other block, labelled by $J\neq I$. To show that $\mathcal{F}_0$ cancels the contribution in equation~\eqref{eq:odd-residual}, we use the Cayley--Hamilton identity for the $2\times2$ blocks of $u$,
\begin{equation}\label{eq:cayley-hamilton}
u_I+u_I^{-1}=\tr(u_I)\,\mathbf{1}_2\,.
\end{equation}
For an odd supermatrix $X$, it implies that the upper-right block of the commutator
with the element $t_0$ is given by
\begin{equation}\label{eq:t0-block}
[t_0,X]_{AS}=\frac{i\kappa}{2}\big(\tr(u_A)\,X_{AS}\,u_S-\tr(u_S)\,u_A\,X_{AS}\big)
=\frac{i\kappa}{2}\big(u_A^{-1}X_{AS}\,u_S-u_A\,X_{AS}\,u_S^{-1}\big)\,,
\end{equation}
and the lower-left block follows by exchanging $A$ and $S$. On the odd subspace we thus have
\begin{equation}\label{eq:t0-odd}
[t_0,X]=\frac{i\kappa}{2}\,\big(\Ad_u^{-1}-\Ad_u\big)X\,.
\end{equation}
Given equation~\eqref{eq:brane-currents}, the contraction of $\mathcal{F}_0$ in the boundary variation is
\begin{equation}\label{eq:F0-contraction}
\mathcal{F}_0(\Delta,\partial_\tau)=\str\big(\Delta_R\,[j_R^\tau,t_0]\big)\,.
\end{equation}
According to equation~\eqref{eq:t0-odd}, its odd part takes the form
\begin{equation}\label{eq:F0-odd}
\str\big(\Delta_R^F\,[j_R^{F,\tau},t_0]\big)
=\frac{i\kappa}{2}\,\str\Big(\Delta_R^F\,\big(\Ad_u-\Ad_u^{-1}\big)\,j_R^{F,\tau}\Big)\,,
\end{equation}
which exactly cancels the contribution in equation~\eqref{eq:odd-residual}. Equivalently, the closed form cancels the fermionic part of the GS term at the boundary,
\begin{equation}\label{eq:odd-cancellation}
\kappa\,\str\big(j_1\wedge j_3\big)+\mathcal{F} =\chi\,\omega+\str\big(t_0\,\theta^B\wedge\theta^B\big)\,.
\end{equation}
All fermionic contributions to the boundary variation therefore cancel, for all values of the fluxes, provided we impose the fermionic boundary conditions in equation~\eqref{eq:fermionic-bc} and choose $\mathcal{F}_0$ as in equations~\eqref{eq:F0a} and~\eqref{eq:F0b}.

Once the brane geometry \eqref{eq:brane} and the resulting fermionic boundary conditions \eqref{eq:fermionic-bc} are fixed, the choice of $\mathcal F_0$ is essentially unique. Indeed, any additional closed two-form should have vanishing contraction with every allowed fermionic variation to preserve these conditions. By closure, it would also have vanishing contraction with the graded brackets of the corresponding odd tangent fields. In the parametrisation \eqref{eq:representative}, these brackets span the bosonic tangent directions on the brane. Therefore, any additional two-form would have to vanish on both the fermionic and the bosonic directions.

It remains to evaluate the bosonic terms. Multiplying the second relation in equation~\eqref{eq:even-tangent} by $u_I$ and using the Cayley--Hamilton identity~\eqref{eq:cayley-hamilton}, one finds on the tangent directions
\begin{equation}\label{eq:tangent-identities}
u_I\,(j_2^\tau)_I=(j_2^\tau)_I\,u_I^{-1}\,,\qquad
\big(1+\Ad_{u_I}\big)(j_2^\tau)_I=\tr(u_I)\,(j_2^\tau)_I\,u_I^{-1}\,.
\end{equation}
With these identities, the bosonic contributions of the kinetic term, of the potential $\chi\,\omega$ and of $\mathcal{F}_0$ all become proportional to the same tangent matrix, and the boundary variation takes the form
\begin{equation}\label{eq:boundary-variation-bosonic}
\delta S\big|_{\partial M}=-\frac{\sqrt\lambda}{2\pi}\int d\tau\,\Big[\str\Big(\Delta_R^B\,\big([u,j_2^\sigma]-\Lambda\,j_2^\tau\big)\,u^{-1}\Big)\Big]_{\sigma=0}^{\sigma=\pi}\,,
\end{equation}
with the constant block-diagonal matrix
\begin{equation}\label{eq:Lambda}
\Lambda=\mathrm{diag}\big(\Lambda_A\mathbf{1}_2,\Lambda_S\mathbf{1}_2\big)\,,\qquad \Lambda_I=\chi\,\tr(u_I)+i\kappa\,\tr(u_J)\,, \qquad J\neq I\,.
\end{equation}
The first term in $\Lambda_I$ arises from the potential $\chi\,\omega$, and the second from the closed form $\mathcal{F}_0$. Since the tangent variations are arbitrary, the boundary variation vanishes if and only if
\begin{equation}\label{eq:polynomial-bc}
\big[u_I,(j_2^\sigma)_I\big]=\Lambda_I\,(j_2^\tau)_I\,.
\end{equation}
When $\tr(u_I)\neq0$, we multiply this condition by $u_I^{-1}$ and use equation~\eqref{eq:tangent-identities}. This gives the tangential gluing condition
\begin{equation}\label{eq:gluing}
(j_2^\sigma)_I=-c_I\,\frac{1+\Ad_{u_I}}{1-\Ad_{u_I}}\,(j_2^\tau)_I\,,\qquad
c_I=\chi+i\kappa\,\frac{\tr(u_J)}{\tr(u_I)}\,, \qquad J\neq I\,.
\end{equation}
Since $\tr(u_A)$ is imaginary and $\tr(u_S)$ is real, see equation~\eqref{eq:u-traces}, the coefficients $c_I$ are real.
\smallskip

The boundary conditions were derived in the representative~\eqref{eq:representative}. The gauge-covariant form is obtained by replacing $u$ with $\hat u$, as defined in equation~\eqref{eq:u-hat}. Collecting our results, the AdS$_2\times$S$^2$ brane with the field strength in equation~\eqref{eq:F-general}, where $\mathcal{F}_0$ and $\omega$ are given by equations~\eqref{eq:F0a}, \eqref{eq:F0b} and~\eqref{eq:omega}, imposes the boundary conditions
\begin{equation}\label{eq:complete-bc}
\begin{aligned}
&(j_2^\sigma)_I=-c_I\,\frac{1+\Ad_{\hat u_I}}{1-\Ad_{\hat u_I}}\,(j_2^\tau)_I &&\text{(tangent)}\,,\\
&(j_2^\tau)_I=0&&\text{(normal)}\,,\\[2mm]
&\big(1-i\Ad_{\hat u}\big)\,j_1^\tau=\big(1+i\Ad_{\hat u}\big)\,j_3^\tau\,,
\end{aligned}
\end{equation}
where the normal directions are those that commute with $\hat u$. For a vanishing trace, the tangential condition is replaced by the condition in equation~\eqref{eq:polynomial-bc}, with $u$ replaced by $\hat u$. Under the twisted diagonal action $(g_L,g_R)\mapsto(w^{-1}hw\,g_L,h\,g_R)$ of the supergroup, the element $\hat u$ is invariant, and so are the boundary conditions. They thus preserve the $PSU(1,1|2)$ subgroup of subsection~\ref{sec:spacetime-branes}.
\smallskip

Let us discuss some limits. For pure NSNS flux, $t_0$ vanishes and $c_I=1$, so that the bosonic conditions reduce to the WZW gluing conditions of equation~\eqref{eq:bosonic-bc-lc} at $\chi=1$. For pure RR flux, on the other hand, the coefficients $c_I$ do not vanish for generic positions of the brane. In contrast to the bosonic condition~\eqref{eq:bosonic-bc} at $\chi=0$, the bosonic gluing conditions are therefore not of Neumann type. Only if both traces vanish, i.e.\ for $\psi=0$ and $\alpha=\pi/2$, does $t_0$ vanish and the condition in equation~\eqref{eq:polynomial-bc} reduces to Neumann conditions along the brane for all values of the fluxes. This is the constant automorphism gluing of the type considered by Dekel and Oz \cite{Dekel:2011ja}. Note that setting the fermions to zero in equation~\eqref{eq:complete-bc} does not take us back to the bosonic condition~\eqref{eq:bosonic-bc}. The presence of fermions and of a non-trivial $\mathcal{F}_0$ changes the dependence of the bosonic gluing condition on the fluxes, replacing $\chi$ by the coefficients $c_I$. As we now show, these coefficients encode the worldvolume fields that we derived in subsection~\ref{sec:spacetime-branes}.

For an open string ending on a brane with worldvolume field $\mathcal{F}$, the tangential gluing condition takes the form
\begin{equation}\label{eq:flat-gluing}
\partial_\sigma X^a=-\mathcal{F}^a{}_b\,\partial_\tau X^b
\end{equation}
in an orthonormal frame on the brane. In the gluing condition~\eqref{eq:gluing}, the role of $\mathcal{F}$ is played by the operator in front of $(j_2^\tau)_I$. On the tangent directions, $\Ad_{u_A}$ has eigenvalues $-e^{\pm2\psi}$ and $\Ad_{u_S}$ has eigenvalues $e^{\pm2i\alpha}$. The operator $(1+\Ad_{u_I})/(1-\Ad_{u_I})$ therefore acts on AdS$_2$ as a boost generator of strength $\tanh\psi$, and on S$^2$ as a rotation generator of strength $\cot\alpha$. Hence the gluing condition describes a brane with the fields
\begin{equation}\label{eq:c-ratio}
\mathsf{e}=c_A\tanh\psi\,,\qquad \mathsf{b}=c_S\cot\alpha\,.
\end{equation}
For pure NSNS flux the coefficients $c_I$ are equal to one, and we recover the fields in equation~\eqref{eq:fields-NSNS}. For general fluxes, the traces in equation~\eqref{eq:u-traces} turn the coefficients in equation~\eqref{eq:gluing} into
\begin{equation}\label{eq:c-match}
c_A=\chi+\kappa\,\frac{\cos\alpha}{\sinh\psi}
=\chi+\kappa\,\frac{\mathsf{h}_0}{\mathsf{e}_0}\,,\qquad
c_S=\chi-\kappa\,\frac{\sinh\psi}{\cos\alpha}
=\chi-\kappa\,\frac{\mathsf{d}_0}{\mathsf{b}_0}\,,
\end{equation}
where the second expressions follow from equations~\eqref{eq:fields-NSNS} and~\eqref{eq:dual-fields-NSNS}. Inserted into equation~\eqref{eq:c-ratio}, these coefficients reproduce precisely the fields in equation~\eqref{eq:rotated-fields}, which we derived from the spacetime dynamics of the brane. The ratios of block traces in the gluing condition are thus the ratios of the dual fields to the original fields of the pure NSNS brane. The fundamental string couples only to $\mathcal{F}$, not to its dual, and therefore sees the rotated combination. Note that the bosonic condition~\eqref{eq:bosonic-bc}, with $\chi$ in place of $c_I$, would describe the fields $\chi\,\mathsf{e}_0$ and $\chi\,\mathsf{b}_0$. These do not solve the spacetime equations of motion~\eqref{eq:brane-eom}, unless $\kappa$ vanishes or the brane sits at $\psi=0$ and $\alpha=\pi/2$. The agreement with the spacetime dynamics of the brane
thus relies on the fermions.

\section{Integrability with boundaries}\label{sec:integrability}

In this section we show that the boundary conditions of section~\ref{sec:branes} preserve the classical integrability of the supercoset sigma model. To this end, we lift them to gluing conditions for the Lax connection~\eqref{eq:lax}. These conditions in turn imply that the double-row monodromy evolves by conjugation, so its spectral invariants are conserved. We first review the bosonic case and then include fermions. Throughout this section we assume $\kappa>0$ and work at generic spectral parameter. The pure NSNS endpoint is a singular limit of the Lax connection parametrisation used here.

\subsection{Review of the bosonic case}\label{sec:bosonic-integrability}

As a warm-up, we consider the purely bosonic case, in which the fermionic currents $j_1$ and $j_3$ vanish. We fix the gauge $g_R=1$ and allow for a nonzero worldvolume flux. This case was discussed thoroughly in reference~\cite{Gil:2026ost}. We review it here for completeness and to clarify some of its geometric aspects. Following the construction presented there, we consider the double-row monodromy matrix,
\begin{equation}\label{eq:double-row}
\mathcal{T}(x)=K_0(x,\tau)\,T^{g_L}(\pi,0;-1/x)^{-1}\,K_\pi(x,\tau)^{-1}\,
T(\pi,0;x)\,,
\end{equation}
where $T$ is the transport matrix in equation~\eqref{eq:transport}, and $T^{g_L}$ is the corresponding transport matrix of the fixed-frame connection $\Lax^{g_L}$, with
\begin{equation}\label{eq:lax-fixed-frame}
\Lax^{g_{L,R}}(x)=g_{L,R}\,\Lax(x)\,g_{L,R}^{-1}-dg_{L,R}\,g_{L,R}^{-1}\,.
\end{equation}
$K_0(x,\tau)$ and $K_\pi(x,\tau)$ are two \textit{reflection matrices} that need to be constructed such that the double-row monodromy matrix~\eqref{eq:double-row} generates infinitely many conserved charges. It is not difficult to see that this is the case if the boundary conditions lift to the Lax gluing conditions in the sense that
\begin{equation}\label{eq:bosonic-lax-gluing}
\Lax^\tau(x)=K_\sigma(x,\tau)\,\Lax^{g_L,\tau}(-1/x)\,K_\sigma^{-1}(x,\tau)
-\partial_\tau K_\sigma(x,\tau)\,K_\sigma^{-1}(x,\tau)\,,\qquad \sigma=0,\pi\,.
\end{equation}
Indeed, substituting these gluing conditions at the two endpoints into equation~\eqref{eq:transport-evolution}
gives
\begin{equation}\label{eq:double-row-evolution}
\partial_\tau\mathcal{T}(x)=\big[\mathcal{T}(x),\Lax^\tau(0;x)\big]\,,
\end{equation}
so that the supertraces of all powers of $\mathcal{T}(x)$ are conserved,
\begin{equation}\label{eq:monodromy-invariants}
Q_n(x)=\str\big(\mathcal T(x)^n\big), \qquad \partial_\tau Q_n(x)=0\,,
\qquad n=1,2,\ldots.
\end{equation}
As usual, to obtain the individual conserved quantities, we expand $Q_n(x)$ in $x$, around a fixed regular point $x_0$. Crucially, in equation~\eqref{eq:bosonic-lax-gluing} the reflection matrices $K_\sigma$ are allowed to depend on $\tau$. Indeed, as pointed out in~\cite{Gil:2026ost}, choosing a constant reflection matrix imposes gluing conditions that select only very special conjugacy classes. More precisely, choosing the constant matrix $K_\sigma=w$, with $w$ given in equation~\eqref{eq:brane-traces},
gives the conditions
\begin{equation}\label{eq:constant-bosonic-gluing}
(1+\Ad_{wg_L})j_2^\tau=0,
\qquad
(1-\Ad_{wg_L})j_2^\sigma=0.
\end{equation}
In the normal directions, the first equation fixes $j_2^\tau=0$, i.e.\ Dirichlet boundary conditions. Recall that, in the parametrisation of equations~\eqref{eq:brane-coordinates} and~\eqref{eq:brane-traces}, $\psi$ and $\alpha$ are identified precisely as the normal directions that label the AdS$_2\times$S$^2$ brane. Expressing the conditions~\eqref{eq:constant-bosonic-gluing} in terms of the coordinates~\eqref{eq:brane-coordinates}, we see that they select a regular conjugacy class with $\psi=0$ and $\alpha=\pi/2$. The values $\alpha=0$ and $\alpha=\pi$ are also allowed, but correspond to collapsed spherical branes.

A nonconstant reflection matrix is therefore necessary to obtain conjugacy classes at generic $\psi$ and $\alpha$, and the corresponding boundary conditions read
\begin{equation}\label{eq:bosonic-integrable-condition}
j_2^\sigma=-\frac{1+\Ad_{wg_L}}{1-\Ad_{wg_L}}\,j_2^\tau\quad\text{(tangent)}\,,
\qquad j_2^\tau=0\quad\text{(normal)}\,.
\end{equation}
As explained in subsection~\ref{sec:boundary-conditions}, this is precisely the condition that follows from the boundary action, with the choice $\mathcal{F}_0=(1-\chi)\,\omega$. The reflection matrix that reproduces this boundary condition was originally found in~\cite{Gil:2026ost} by means of an ansatz in terms of the generators of the supergroup. Here we provide a more conceptual derivation and clarify the geometric interpretation of this result. The starting point is the observation that a brane at generic values of $\psi$ and $\alpha$ can be obtained from the special configuration at $\psi=0$ and $\alpha=\pi/2$ by moving all points of the brane's worldvolume along geodesics normal to the brane. In the
left-invariant frame, these geodesics are generated by the normal currents
\begin{equation}\label{eq:normal-currents}
\begin{aligned}
N_\psi&=g_A^{-1}\partial_\psi g_A
=\cosh\rho\,\sin t\,\sigma_1+\cosh\rho\,\cos t\,\sigma_3+i\sinh\rho\,\sigma_2\,,\\[2mm]
N_\alpha&=g_S^{-1}\partial_\alpha g_S
=i\big(-\sin\beta\sin\gamma\,\sigma_1+\sin\beta\cos\gamma\,\sigma_2+\cos\beta\,\sigma_3\big)\,.
\end{aligned}
\end{equation}
These normal currents act on the AdS and the sphere block, respectively, and they are normalised such that
\begin{equation}\label{eq:N2}
    N^2_\psi=\mathbf{1}_2\,, \qquad N^2_\alpha=-\mathbf{1}_2\,.
\end{equation}
In expressions involving $4\times4$ matrices, we understand $N_\psi$ and $N_\alpha$ as their block embeddings $\mathrm{diag}(N_\psi,0_2)$ and $\mathrm{diag}(0_2,N_\alpha)$, respectively.
The geometric intuition leads us to generalise $K_\sigma=w$ by adding a linear combination of the normal currents,
\begin{equation}\label{eq:normal-reflection}
K_\sigma=\big(\mathbf{1}_4+a_\psi(x,\psi)\,N_\psi+a_\alpha(x,\alpha)\,N_\alpha\big)\,w\,,
\end{equation}
where we suppressed the dependence on $\tau$. As is clear from equation~\eqref{eq:normal-currents}, the normal currents depend on $\tau$ and $\sigma$ through their dependence on the brane's worldvolume coordinates $(\rho,t)$ and $(\beta,\gamma)$. Note that they do not depend on the position parameters $\psi$ and $\alpha$. The dependence on these normal coordinates is encoded in the coefficient functions $a_\psi$ and $a_\alpha$. For a homogeneous brane geometry, these coefficients are expected to be functions of $\psi$ and $\alpha$ only, and hence independent of $\tau$, 
\begin{equation}
    \partial_\tau a_\psi(x,\psi)\;=\;0, \qquad \partial_\tau a_\alpha(x,\alpha)\;=\;0.
\end{equation}
Indeed the normal coordinates $\psi$ and $\alpha$ are independent of $\tau$, since they are fixed by
Dirichlet boundary conditions and label the brane.

The coefficients $a_\psi$ and $a_\alpha$ are determined algebraically by substituting equation~\eqref{eq:normal-reflection} into equation~\eqref{eq:bosonic-lax-gluing} and imposing the boundary condition~\eqref{eq:bosonic-integrable-condition}. Their explicit expression reads
\begin{equation}\label{eq:normal-reflection-coefficients}
\begin{gathered}
a_\psi(x,\psi)=-\frac{\sinh2\psi}{z(x)+\cosh2\psi}\,,\qquad
a_\alpha(x,\alpha)=\frac{\sin2\alpha}{z(x)-\cos2\alpha}\,,\\[2mm]
z(x)=\frac{a(x)+b(x)-1}{1+a(x)-b(x)}
=\frac{(x-1)\big[(\chi-1)(x+1)+\kappa(x-1)\big]}
{(x+1)\big[(1-\chi)(x-1)+\kappa(x+1)\big]}\,.
\end{gathered}
\end{equation}
At $\psi=0$ and $\alpha=\pi/2$, both $a_\psi$ and $a_\alpha$ vanish at generic spectral parameter and one recovers $K_\sigma=w$. One can check that the reflection matrix in equation~\eqref{eq:normal-reflection} indeed reproduces the corresponding result in~\cite{Gil:2026ost}.

Finally, we note that the bosonic\footnote{Once fermions are included, the odd blocks couple the AdS and sphere sectors, so their relative normalisation can no longer be chosen independently.} Lax gluing condition~\eqref{eq:bosonic-lax-gluing} is insensitive to independent constant normalisations of the AdS and sphere blocks of $K_\sigma$. Therefore, we could also rescale the reflection matrix and write it as
\begin{equation}\label{eq:bosonic-exponential-rescaling}
\begin{aligned}
\widetilde K_\sigma
&=\mathrm{diag}\!\left(
\frac{\mathbf1_2}{\sqrt{1-a_\psi^2}},
\frac{\mathbf1_2}{\sqrt{1+a_\alpha^2}}
\right)K_\sigma
=e^{\eta^B_\psi N_\psi}e^{\eta^B_\alpha N_\alpha}w,\\
\eta^B_\psi&=\operatorname{arctanh}a_\psi,\qquad
\eta^B_\alpha=\arctan a_\alpha.
\end{aligned}
\end{equation}
where we used equation~\eqref{eq:N2} to pass to exponentials. In this form, it is apparent that the factor multiplying $w$ represents finite displacements along the normal directions. Indeed, since the currents $N_\psi$ and $N_\alpha$ do not depend on the corresponding normal coordinates, right multiplication by their exponentials shifts $\psi$ by $\eta^B_\psi$ and $\alpha$ by $\eta^B_\alpha$.

\subsection{Integrable boundary conditions including fermions}\label{sec:fermionic-integrability}
We now extend the discussion of the previous subsection to include fermions.
We impose a Lax gluing condition with the same structure as the one in equation~\eqref{eq:bosonic-lax-gluing}, now using both fixed frames,
\begin{equation}\label{eq:full-lax-gluing}
\Lax^{g_R,\tau}(x)=K_\sigma(x,\tau)\,\Lax^{g_L,\tau}(-1/x)\,K_\sigma(x,\tau)^{-1}
-\partial_\tau K_\sigma(x,\tau)\,K_\sigma(x,\tau)^{-1}\,.
\end{equation}
The matrices $g_L$ and $g_R$ are the full supergroup fields, and in the double-row monodromy matrix~\eqref{eq:double-row} the transport matrix $T$ is replaced by the transport matrix $T^{g_R}$ of the connection $\Lax^{g_R}$, i.e.\ we define it as
\begin{equation}\label{eq:double-row-2}
\mathcal{T}(x)=K_0(x,\tau)\,T^{g_L}(\pi,0;-1/x)^{-1}\,K_\pi(x,\tau)^{-1}\,
T^{g_R}(\pi,0;x)\,.
\end{equation}
Note that the gluing condition~\eqref{eq:full-lax-gluing} involves the same reflection of the spectral parameter, $x\mapsto -1/x$, as in the bosonic case. This time, however, it also acts on the square-root\footnote{Changing the branch of the square root in equation~\eqref{eq:lax-coeff} flips the signs of $c_1$ and $c_3$, which can be compensated by changing the signs of $j_1$ and $j_3$. The sign in the lift~\eqref{eq:spectral-lift}, on the other hand, matters. The opposite sign describes the brane with the opposite orientation, i.e.\ it leads to the reflection matrix with $\kappa$ replaced by $-\kappa$ and to the fermionic boundary condition~\eqref{eq:fermionic-bc} with $\Ad_{\hat u}$ replaced by $-\Ad_{\hat u}$.} factors appearing in the fermionic coefficients of the Lax connection~\eqref{eq:lax},
\begin{equation}\label{eq:spectral-lift}
x\longmapsto-\frac1x,\qquad
\sqrt{x^2-1}\longmapsto\frac{i}{x}\sqrt{x^2-1}.
\end{equation}
Applying this map twice gives the identity. Direct substitution in equation~\eqref{eq:lax-coeff} gives the following transformations for the coefficients of the Lax connection,
\begin{equation}\label{eq:full-spectral-reflection}
\begin{aligned}
a(-1/x)&=-a(x),& b(-1/x)&=b(x),\\[2mm]
c_1(-1/x)&=-ic_3(x),& c_3(-1/x)&=ic_1(x).
\end{aligned}
\end{equation}
The relations for $a$ and $b$ hold for either branch of the square root, while those for $c_1$ and $c_3$ rely on the lift~\eqref{eq:spectral-lift}.

We consider the branes and closed two-form introduced in subsection~\ref{sec:boundary-conditions}. Our aim is to show that the corresponding current boundary conditions~\eqref{eq:complete-bc} are equivalent to the Lax gluing condition~\eqref{eq:full-lax-gluing}, for a suitable choice of reflection matrix on the prescribed brane. We shall allow the two endpoints of the open strings to end on two different branes. Therefore the reflection matrices $K_0$ and $K_\pi$ may be constructed with different elements $u$ corresponding to different conjugacy classes. 

We construct the reflection matrix using normal currents, following the same geometric intuition as in the purely bosonic case. To define a normal derivative, we temporarily consider the family of neighbouring branes in the representative~\eqref{eq:representative},
\begin{equation}\label{eq:normal-class-family}
(g_L,g_R)=(w^{-1}su,s)\,, \quad
g(\psi,\alpha;s)=w^{-1}s\,u(\psi,\alpha)s^{-1}\,,\quad
u(\psi,\alpha)=\mathrm{diag}\!\left(
i\sigma_3e^{\psi\sigma_3},e^{i\alpha\sigma_3}\right)\,.
\end{equation}
Here we chose the bosonic base point where the four worldvolume coordinates in equation~\eqref{eq:brane-coordinates} vanish. The full supergroup element $s$ then moves the base point along the superbrane, as in equation~\eqref{eq:representative}. Holding $s$ fixed while varying $\psi$ or $\alpha$ follows the normal geodesic through each such point. The normal currents are therefore defined exactly as before,
\begin{equation}\label{eq:superbrane-normal-generators}
\begin{aligned}
N_\psi&=g^{-1}\partial_\psi g
=s\,\mathrm{diag}(n_\psi,0_2)s^{-1},
&n_\psi&=u_A^{-1}\partial_\psi u_A=\sigma_3,\\[2mm]
N_\alpha&=g^{-1}\partial_\alpha g
=s\,\mathrm{diag}(0_2,n_\alpha)s^{-1},
&n_\alpha&=u_S^{-1}\partial_\alpha u_S=i\sigma_3.
\end{aligned}
\end{equation}
The derivatives hold $s$ fixed. The matrices $n_\psi,n_\alpha$ are simply these currents at the bosonic base point, before transport by $s$. The full currents $N_\psi,N_\alpha$ commute and are independent of $\psi,\alpha$. On the bosonic body they reduce to the normal currents of equation~\eqref{eq:normal-currents}; transport by the full $s$ encodes the fermionic dependence.

We use the same exponential ansatz as in the bosonic case, with two $\tau$-independent displacement coefficients:
\begin{equation}\label{eq:full-reflection}
\begin{aligned}
K_\sigma(x,\tau)
&=e^{\eta_\psi(x)N_\psi}e^{\eta_\alpha(x)N_\alpha}w
=g_R \mathcal R(x)g_L^{-1},\\[2mm]
\mathcal R(x)&=\mathrm{diag}\!\left(
u_Ae^{\eta_\psi(x)n_\psi},u_Se^{\eta_\alpha(x)n_\alpha}\right).
\end{aligned}
\end{equation}
We have used the representative~\eqref{eq:representative}, with the original brane labels fixed as in equation~\eqref{eq:u-traces}. Since $u$ and the displacement coefficients are constant along the boundary, $\mathcal R$ does not depend on $\tau$. Nevertheless, the reflection matrix $K_\sigma$ is generally time dependent and has fermionic entries, since it contains $g_R$ and $g_L^{-1}$. Expanding the exponentials gives
\begin{equation}\label{eq:displaced-exponential-blocks}
\begin{aligned}
\mathcal R_A&=i\sinh(\psi+\eta_\psi)\,\mathbf1_2
+i\cosh(\psi+\eta_\psi)\,n_\psi,\\[2mm]
\mathcal R_S&=\cos(\alpha+\eta_\alpha)\,\mathbf1_2
+\sin(\alpha+\eta_\alpha)\,n_\alpha.
\end{aligned}
\end{equation}
Both blocks have determinant one. Removing the two fixed-frame transformations~\eqref{eq:lax-fixed-frame} in equation~\eqref{eq:full-lax-gluing} cancels the frame derivatives and gives the condition
\begin{equation}\label{eq:lax-intertwining}
\Lax^\tau(x)\mathcal R(x)-\mathcal R(x)\Lax^\tau(-1/x)=0.
\end{equation}
Using the coefficient reflection~\eqref{eq:full-spectral-reflection}, this condition takes the form
\begin{equation}\label{eq:full-component-gluing}
\begin{aligned}
0={}&[j_0^\tau,\mathcal R]+a\{j_2^\tau,\mathcal R\}+b[j_2^\sigma,\mathcal R]\\[2mm]
&+c_1\big(j_1^\tau \mathcal R-i\mathcal Rj_3^\tau\big)
+c_3\big(j_3^\tau \mathcal R+i\mathcal Rj_1^\tau\big).
\end{aligned}
\end{equation}
Its first line is the even part and its second line the odd part of the condition. Braces denote the anticommutator, namely $\{U,V\}=UV+VU$. From here on, all Lax coefficients and reflection matrices in algebraic identities are evaluated at $x$.
We first determine the two displacement coefficients $\eta_\psi(x)$ and $\eta_\alpha (x)$ from the even equation, then check that the resulting matrix gives both boundary conditions.

The constant representative and the current definitions~\eqref{eq:currents} give
\begin{equation}\label{eq:u-constant}
g_R^{-1}wg_L=u,\qquad 0=\partial_\tau u=-j_R^\tau u+u j_L^\tau.
\end{equation}
Taking the even part and using the tangency condition~\eqref{eq:tangent-identities}, we find
\begin{equation}\label{eq:constant-u-even}
[(j_0^\tau)_I,u_I]=\{u_I,(j_2^\tau)_I\}=\tr(u_I)(j_2^\tau)_I.
\end{equation}
The bosonic boundary condition~\eqref{eq:polynomial-bc}, written with the normals~\eqref{eq:superbrane-normal-generators}, reduces to
\begin{equation}\label{eq:normal-field-gluing}
\begin{aligned}
(j_2^\sigma)_A&=\mathsf e\,n_\psi(j_2^\tau)_A,
&\mathsf e&=\frac{\chi\sinh\psi+\kappa\cos\alpha}{\cosh\psi},\\[2mm]
(j_2^\sigma)_S&=-\mathsf b\,n_\alpha(j_2^\tau)_S,
&\mathsf b&=\frac{\chi\cos\alpha-\kappa\sinh\psi}{\sin\alpha},
\end{aligned}
\qquad\text{(tangent)}.
\end{equation}
Recall that these conditions include the contribution of $\mathcal F_0$, as given in equations~\eqref{eq:F0a} and~\eqref{eq:F0b}. We use them to fix the displacement coefficients: substituting equations~\eqref{eq:displaced-exponential-blocks}, \eqref{eq:constant-u-even} and~\eqref{eq:normal-field-gluing} into the even part of equation~\eqref{eq:full-component-gluing} gives
\begin{equation}\label{eq:even-lax-terms}
\begin{aligned}
0&=2i\cosh(\psi+\eta_\psi)
\big[\tanh\psi+a\tanh(\psi+\eta_\psi)-b\, \mathsf e\big](j_2^\tau)_A,\\[2mm]
0&=2\sin(\alpha+\eta_\alpha)
\big[\cot\alpha+a\cot(\alpha+\eta_\alpha)-b\, \mathsf b\big](j_2^\tau)_S.
\end{aligned}
\end{equation}
Here the tangent currents anticommute with the corresponding normals, as follows from equation~\eqref{eq:tangent-identities}; normal components of $j_2^\sigma$ commute with $\mathcal R$ and do not contribute. For arbitrary tangent motion, the two bracketed coefficients must vanish. Solving these scalar equations gives
\begin{equation}\label{eq:full-normal-coefficients}
\begin{aligned}
\tanh\eta_\psi &=\frac{b\, \mathsf e-(1+a)\tanh\psi}
{a-b\, \mathsf e\tanh\psi+\tanh^2\psi},\\[2mm]
\tan\eta_\alpha &=\frac{(1+a)\cot\alpha-b\, \mathsf b}
{a+b\, \mathsf b\cot\alpha-\cot^2\alpha}.
\end{aligned}
\end{equation} 
For $c_A=c_S=1$, i.e.\ for the fields $\mathsf{e}_0$ and $\mathsf{b}_0$ of the pure NSNS brane in
equation~\eqref{eq:fields-NSNS}, these relations reduce to
\begin{equation}\label{eq:displacements-bosonic-limit}
\tanh\eta_\psi=a_\psi\,,\qquad \tan\eta_\alpha=a_\alpha\,,
\end{equation}
with the coefficients in equation~\eqref{eq:normal-reflection-coefficients}. In this case the 
displacements coincide with those of the bosonic construction in subsection~\ref{sec:bosonic-integrability}.  The closed two-form $\mathcal{F}_0$ that is required by the fermions affects the displacements only through  the fields $\mathsf{e}$ and $\mathsf{b}$.  Going back to the general case the even equation fixes each block separately. To obtain a reflection of the  full supermatrix, we must also fix their relative sign, since the presence of odd elements entangles the AdS  and sphere blocks. Using the vanishing brackets in equation~\eqref{eq:even-lax-terms}, the 
fields~\eqref{eq:normal-field-gluing} and the bulk relation in equation~\eqref{eq:coefficient-identities}, we obtain
\begin{equation}\label{eq:normal-displacement-identity}
\begin{aligned}
\frac{\cosh^2\psi}{\cosh^2(\psi+\eta_\psi)}
-\frac{\sin^2\alpha}{\sin^2(\alpha+\eta_\alpha)}
&=\frac{\sinh^2\psi+\cos^2\alpha}{a^2} \big(a^2-b^2+2\chi b-1\big)=0.
\end{aligned}
\end{equation} 
For real spectral parameter, the second relation in equation~\eqref{eq:even-lax-terms} has a solution for which $\alpha+\eta_\alpha$ lies between zero and $\pi$. Equation~\eqref{eq:normal-displacement-identity} then implies that the magnitude of $\tanh(\psi+\eta_\psi)$ is smaller than one, so that the first relation admits a real displacement $\eta_\psi$. With these branches, equation~\eqref{eq:normal-displacement-identity} gives
\begin{equation}\label{eq:linked-normal-displacements}
\frac{\cosh(\psi+\eta_\psi)}{\cosh\psi} =\frac{\sin(\alpha+\eta_\alpha)}{\sin\alpha}>0.
\end{equation}
For complex spectral values we continue these branches locally. The relative sign is now fixed. Combining this relation with equations~\eqref{eq:displaced-exponential-blocks} and~\eqref{eq:even-lax-terms} gives the explicit reflection matrix
\begin{equation}\label{eq:displaced-reflection-blocks}
\begin{gathered}
\mathcal R_I=\frac{\cosh(\psi+\eta_\psi)}{\cosh\psi}
\left[u_I-\frac{(1+a-\chi b)\tr(u_I)-i\kappa b\tr(u_J)}{2a}\,\mathbf1_2
\right],\\
(I,J)=(A,S)\ \text{or}\ (S,A).
\end{gathered}
\end{equation}
By equation~\eqref{eq:linked-normal-displacements}, the prefactor is the same for the two blocks.

We now verify that the Lax gluing condition~\eqref{eq:full-lax-gluing} with this $\mathcal R$ implies both the even and odd boundary conditions. We start from the even part, which should trivially hold, given the way we derived $\mathcal R$. Substitution in the even part of equation~\eqref{eq:full-component-gluing}, together with equation~\eqref{eq:constant-u-even}, gives
\begin{equation}\label{eq:even-lax-residual}
0=\frac{b\cosh(\psi+\eta_\psi)}{\cosh\psi}
\left\{\big(\chi\tr(u_I)+i\kappa\tr(u_J)\big)(j_2^\tau)_I
-[u_I,(j_2^\sigma)_I]\right\}.
\end{equation}
Clearly, this equation holds if and only if the bosonic boundary condition~\eqref{eq:polynomial-bc} holds.

It remains to check that the same $\mathcal R$ works for the fermions. The current condition obtained in subsection~\ref{sec:boundary-conditions} is
\begin{equation}\label{eq:odd-current-target}
j_R^{F,\tau}=\Ad_u j_L^{F,\tau}.
\end{equation}
Expressing the odd part of equation~\eqref{eq:full-component-gluing} in terms of these currents, using equation~\eqref{eq:Z4-components} and multiplying on the right by $2\mathcal R^{-1}$, gives
\begin{equation}\label{eq:odd-lax-expanded}
\begin{aligned}
&(c_1+c_3)j_L^{F,\tau}+i(c_1-c_3)j_R^{F,\tau}\\[2mm]
&\qquad=\Ad_{\mathcal R}\big[i(c_1-c_3)j_L^{F,\tau}+(c_1+c_3)j_R^{F,\tau}\big].
\end{aligned}
\end{equation}
To make this equation reduce to the desired condition, $\mathcal R$ must relate the two weighted combinations of the odd currents. We check this directly from its blocks. The coefficient identities~\eqref{eq:coefficient-identities} imply
\begin{equation}\label{eq:odd-coefficient-identities}
\begin{aligned}
(1-a-\chi b)(c_1+c_3)-\kappa b(c_1-c_3)&=0,\\[2mm]
(1+a-\chi b)(c_1-c_3)+\kappa b(c_1+c_3)&=0.
\end{aligned}
\end{equation}
For an arbitrary odd block $X$, substitute equation~\eqref{eq:displaced-reflection-blocks} and reduce the squares of $u_I$ using the Cayley--Hamilton identity~\eqref{eq:cayley-hamilton}. The mixed products cancel because the coefficient of $u_I$ is common to both blocks, and the remaining terms vanish by equation~\eqref{eq:odd-coefficient-identities}. The resulting relation is
\begin{equation}\label{eq:odd-block-intertwiner}
\begin{aligned}
&\mathcal R_I\big[i(c_1-c_3)Xu_J+(c_1+c_3)u_IX\big]\\[2mm]
&\qquad=\big[(c_1+c_3)Xu_J+i(c_1-c_3)u_IX\big]\mathcal R_J.
\end{aligned}
\end{equation}
Each block of $\mathcal R$ commutes with the corresponding block of $u$. Multiplying on the right by $u_J^{-1}\mathcal R_J^{-1}$ therefore gives the identity on odd matrices that we need:
\begin{equation}\label{eq:odd-operator-identity}
\begin{aligned}
\Ad_{\mathcal R}\big[i(c_1-c_3)\mathrm{id}_F+(c_1+c_3)\Ad_u\big]
={}&(c_1+c_3)\mathrm{id}_F\\[2mm]
&+i(c_1-c_3)\Ad_u.
\end{aligned}
\end{equation}
Here $\mathrm{id}_F$ denotes the identity on odd matrices. Using this identity in equation~\eqref{eq:odd-lax-expanded} factors the odd Lax equation as
\begin{equation}\label{eq:odd-lax-factored}
\big[i(c_1-c_3)\mathrm{id}_F-(c_1+c_3)\Ad_{\mathcal R}\big]
\big(j_R^{F,\tau}-\Ad_u j_L^{F,\tau}\big)=0.
\end{equation}
The second factor is exactly the fermionic boundary condition. To see that it is also necessary, multiply the operator in front by the one appearing in equation~\eqref{eq:odd-operator-identity}. That identity and the sum of the first two bulk relations in equation~\eqref{eq:coefficient-identities} give
\begin{equation}\label{eq:odd-lax-inverse}
\begin{aligned}
&\big[i(c_1-c_3)\mathrm{id}_F-(c_1+c_3)\Ad_{\mathcal R}\big]
\circ\big[i(c_1-c_3)\mathrm{id}_F+(c_1+c_3)\Ad_u\big]\\[2mm]
&\qquad=-2(c_1^2+c_3^2)\mathrm{id}_F=-4a\,\mathrm{id}_F.
\end{aligned}
\end{equation}
The prefactor in equation~\eqref{eq:odd-lax-factored} is therefore invertible at generic values where $a$ is nonzero. The odd Lax equation is equivalent to the fermionic current condition, with no additional constraint.

We have thus constructed a reflection matrix whose Lax gluing imposes exactly the expected current boundary conditions. The gluing conditions imply that the double-row monodromy~\eqref{eq:double-row-2} evolves by conjugation and generates infinitely many conserved charges, as in equation~\eqref{eq:double-row-evolution} with $\Lax^\tau$ replaced by $\Lax^{g_R,\tau}$, so that the invariants in equation~\eqref{eq:monodromy-invariants} are conserved. The calculations in this section were made in the representative with constant $u$. Under a bosonic gauge transformation, $\mathcal R$ transforms by conjugation and $u$ is replaced by the element $\hat u$ in equation~\eqref{eq:u-hat}, while $K_\sigma$ and the two fixed-frame connections are unchanged. The gluing condition in the current frame then includes the derivative of $\mathcal R$ with respect to $\tau$, which is exactly the term needed for covariance. The computation can be repeated in this case and one obtains the same conditions derived above, with $\Ad_u$ replaced by $\Ad_{\hat u}$, corresponding to the boundary conditions~\eqref{eq:complete-bc}.
\smallskip

\section{Conclusions}\label{sec:conclusions}

In this work we have closed a gap in the analysis of integrable branes in AdS backgrounds by Dekel and Oz \cite{Dekel:2011ja}, which did not address AdS$_3$ backgrounds with mixed fluxes. More specifically, we have described a family of half-BPS AdS$_2\times$S$^2$ branes in type IIB superstring theory on AdS$_3\times$S$^3\times$T$^4$ with mixed NSNS and RR fluxes. From the spacetime perspective, these are D3-branes that sit at a point of the torus, and we determined their worldvolume fields $\mathsf{e}$ and $\mathsf{b}$ for all values of the fluxes and all positions of the branes, see equation~\eqref{eq:rotated-fields}. They are obtained from the fields of the familiar AdS$_2$ and S$^2$ branes of the WZW model through the same duality rotation that relates the mixed flux background to the pure NSNS one. From the worldsheet perspective, we derived the gluing conditions that the currents of the Green--Schwarz superstring obey at the ends of an open string, see equation~\eqref{eq:complete-bc}. A crucial ingredient is the closed two-form $\mathcal{F}_0$ in the boundary action, which cancels the fermionic boundary terms of the GS term and, at the same time, modifies the bosonic gluing condition. In particular, the bosonic currents obey deformed WZW gluing conditions, while the branes remain twisted conjugacy classes\footnote{A related phenomenon occurs in $\lambda$-deformed WZW models, where the deformed gluing conditions follow from requiring integrability \cite{Driezen:2018glg}. We thank Sibylle Driezen for pointing out this connection.}. Remarkably, the gluing conditions reproduce precisely the worldvolume fields that we found from the spacetime perspective. We then showed that the open Green--Schwarz superstring with these boundary conditions is classically integrable, extending the analysis of \cite{Gil:2026ost} to include fermions. In particular, we constructed the double-row monodromy matrix in equation~\eqref{eq:double-row-2}, with the reflection matrices defined in equation~\eqref{eq:full-reflection}, and showed that it evolves by conjugation. Its eigenvalues, which depend on the spectral parameter, are therefore time independent. Remarkably, the reflection matrices admit a simple geometric interpretation in terms of displacements along geodesics normal to the brane. These results open the door for further studies.

At the classical level, the double-row monodromy matrix that we constructed is the natural starting point for a boundary version of the algebraic curve, which would extend the finite-gap equations for closed strings in the mixed flux background~\cite{Babichenko:2014yaa} to open strings. At the quantum level, one would like to determine the reflection matrices of the worldsheet excitations and, ultimately, the exact spectrum of open strings that end on the branes. The prospects for such a programme depend on the status of the bulk theory, which differs considerably between pure RR and mixed fluxes, see~\cite{Sfondrini:2014via,Demulder:2023bux,Seibold:2024qkh} for reviews. In this literature, the fraction $\chi$ of NSNS flux is usually denoted by $q$. For pure RR flux, where this programme was initiated~\cite{Babichenko:2009dk,OhlssonSax:2011ms}, the spectral problem is now under good control. The worldsheet S matrix including the massless modes~\cite{Borsato:2013qpa,Borsato:2014hja} and its dressing factors are known, and they give rise to a mirror thermodynamic Bethe ansatz for the spectrum~\cite{Frolov:2021bwp,Frolov:2021zyc,Frolov:2021fmj}, whose tensionless limit is also understood~\cite{Brollo:2023pkl,Brollo:2023rgp}. The quantum spectral curve, which was proposed in~\cite{Ekhammar:2021pys,Cavaglia:2021eqr} and studied further in~\cite{Cavaglia:2022xld,Ekhammar:2024kzp}, has recently been derived from the mirror thermodynamic Bethe ansatz~\cite{Cavaglia:2026scj,Cavaglia:2026tqc}, and it has been solved numerically from weak to strong coupling~\cite{Ekhammar:2026ykk}.

For mixed fluxes, the quantum theory is considerably less developed. The worldsheet S matrix is fixed by symmetry up to dressing factors~\cite{Lloyd:2014bsa}, but the worldsheet kinematics and the crossing equations have been analysed only recently~\cite{Frolov:2023lwd,OhlssonSax:2023qrk}. The dressing factors that were proposed in~\cite{OhlssonSax:2023qrk,Frolov:2024pkz,Frolov:2025uwz} have led to a mirror thermodynamic Bethe ansatz~\cite{Frolov:2025tda}, and it remains to be seen whether the quantum spectral curve can be extended to mixed fluxes. At the pure NSNS point, finally, the worldsheet theory in light-cone gauge can be understood as an integrable deformation of a free theory~\cite{Baggio:2018gct}. For open strings, much less is known. Reflection matrices of the worldsheet excitations have so far been studied mostly for branes in backgrounds with pure RR flux~\cite{Prinsloo:2015apa,Bielli:2024xuv}. The branes we considered provide integrable boundaries for all values of the fluxes, and it would be very interesting to determine their reflection matrices. Since the classical reflection matrices depend on the boundary values of the fields, as for some branes in AdS$_5\times$S$^5$~\cite{Linardopoulos:2021rfq,Demjaha:2025axy}, it will also be important to understand how this dependence manifests itself in the quantum theory.

Let us finally come back to holography. The branes we studied are expected to possess a dual description as interfaces in the two-dimensional conformal field theory that is dual to string theory on AdS$_3\times$S$^3\times$T$^4$. This conformal field theory is a deformation of the symmetric product orbifold of T$^4$. Boundaries and interfaces in symmetric product orbifolds were recently studied by several groups \cite{Belin:2021nck,Gaberdiel:2021kkp,Baig:2023ahz,Harris:2025wak,Harris:2025klp,Belleri:2025eun}, see also \cite{Gutperle:2024rvo, Gutperle:2024vyp, Knighton:2024noc,Ghasemi:2026sij} for topological interfaces. A particularly precise version of the correspondence arises for one unit of NSNS flux, where type IIB superstring theory on AdS$_3\times$S$^3\times$T$^4$ is dual to the symmetric product orbifold of T$^4$ at its free point \cite{Gaberdiel:2018rqv,Eberhardt:2018ouy,Eberhardt:2019ywk}. In this setting, D-branes were matched with boundary states of the symmetric product orbifold in reference~\cite{Gaberdiel:2021kkp}. Moreover, it was shown in reference~\cite{Harris:2025wak} that the spectrum of interface changing operators for a family of interfaces in the symmetric product orbifold reproduces the open string partition function of AdS$_2\times$S$^2$ branes, see also~\cite{Harris:2025klp}. Although our analysis applies to the classical worldsheet theory at large string tension, far from the free point, it suggests that these interfaces possess an integrable deformation away from the free orbifold theory, namely the deformation that corresponds to switching on RR flux in the dual geometry. For the bulk theory, this deformation of the symmetric product orbifold has been studied in a number of papers \cite{Pakman:2009mi,Gaberdiel:2015uca,Apolo:2022fya,Fiset:2022erp,Gaberdiel:2023lco,Frolov:2023pjw,Fabri:2025rok,Gaberdiel:2025smz}. In particular, the perturbation by a marginal operator from the twist-two sector gives rise to anomalous dimensions that appear to be governed by an integrable S matrix \cite{Gaberdiel:2023lco}, which maps to the small-flux limit of the S matrix that was bootstrapped from worldsheet integrability \cite{Frolov:2023pjw}. It would be interesting to address the corresponding deformation of the interface theories and, potentially, to construct the all-loop spectrum of interface changing operators, which might turn out to be simpler than in the case of the bulk theory.

\acknowledgments
We would like to thank Bogdan Stefanski for suggesting this problem to us and Julio Cabello Gil, Sibylle Driezen, Sebastian Harris, Torben Skrzypek for discussions. This research is funded by the Deutsche Forschungsgemeinschaft (DFG, German Research Foundation) under the Collaborative Research Center -- SFB 1624 ``Higher structures, moduli spaces and integrability'' -- 506632645. It is also supported by the Deutsche Forschungsgemeinschaft under Germany's Excellence Strategy -- EXC 2121 ``Quantum Universe'' -- 390833306. We used the language models Claude (Anthropic) and ChatGPT (OpenAI) to assist with polishing the text and checking computations. The authors take full responsibility for the content of this paper.
\appendix

\section{The two-form potential on the brane}\label{app:potential}

In this appendix we prove the expression~\eqref{eq:H-g} for the three-form $H$ and the last relation in equation~\eqref{eq:omega}, which states that $\omega$ is a potential for the pullback of $H$ to the brane. Throughout the appendix, powers of one-forms denote wedge products.

We start from the definition~\eqref{eq:H},
\begin{equation}\label{eq:H-app}
H=2\str\Big(\tfrac23\,j_2\wedge j_2\wedge j_2 +j_1\wedge j_3\wedge j_2+j_3\wedge j_1\wedge j_2\Big)\,,
\end{equation}
and express it in terms of $j_L$ and $j_R$. Using the decomposition~\eqref{eq:Z4-inverse} and the cyclicity of the supertrace, the first cubic term becomes
\begin{equation}\label{eq:app-cubic}
\begin{aligned}
\tfrac{4}{3}\str\big(j_2^3\big)
={}&\tfrac{2}{3}\str\big(j_L^3\big)-\tfrac{2}{3}\str\big(j_R^3\big)
-4\str\big(j_0\wedge j_0\wedge j_2\big)\\
&-4\str\big(j_0\wedge(j_1\wedge j_1+j_3\wedge j_3)\big)\\
&-4\str\big(j_2\wedge(j_1\wedge j_3+j_3\wedge j_1)\big)\,.
\end{aligned}
\end{equation}
Here we have dropped terms that contain an odd number of fermionic currents, since their supertrace vanishes. For the mixed cubic terms, using the Maurer--Cartan equations~\eqref{eq:MC-Z4} and the same supertrace properties, we find
\begin{equation}\label{eq:app-exact}
\begin{aligned}
2\str\big((j_1\wedge j_3+j_3\wedge j_1)\wedge j_2\big)
={}&-2\str\big(j_2^3\big)-2\,d\,\str\big(j_0\wedge j_2\big)\\
&+2\str\big(j_0\wedge j_0\wedge j_2\big)\\
&+2\str\big(j_0\wedge(j_1\wedge j_1+j_3\wedge j_3)\big)\,.
\end{aligned}
\end{equation}
Combining equations~\eqref{eq:app-cubic} and~\eqref{eq:app-exact}, we obtain
\begin{equation}\label{eq:app-H}
H=\frac16\Big(\str\big(j_L^3\big)-\str\big(j_R^3\big)\Big)
-d\,\str\big(j_0\wedge j_2\big)\,.
\end{equation}
Next we express the cubic terms through the field $g$ in equation~\eqref{eq:g}. Using the second relation in equation~\eqref{eq:g}, the Maurer--Cartan equations~\eqref{eq:MC-LR} and cyclicity, one obtains
\begin{equation}\label{eq:app-relative}
\str\big((g^{-1}dg)^3\big)=\str\big(j_L^3\big)-\str\big(j_R^3\big) +3\,d\,\str\big(j_L\wedge j_R\big)\,.
\end{equation}
Moreover, the decomposition~\eqref{eq:Z4-inverse} gives
\begin{equation}\label{eq:app-fermionic}
\str\big(j_L\wedge j_R\big)+2\str\big(j_0\wedge j_2\big)
=\str\big(j_L^F\wedge j_R^F\big)=2i\,\str\big(j_1\wedge j_3\big)\,.
\end{equation}
Inserting equations~\eqref{eq:app-relative} and~\eqref{eq:app-fermionic} into equation~\eqref{eq:app-H} gives
\begin{equation}
    H=\frac16\,\str\big(g^{-1}dg\wedge g^{-1}dg\wedge g^{-1}dg\big)
-i\,d\,\str\big(j_1\wedge j_3\big)\,,
\end{equation}
which is precisely the expression in equation~\eqref{eq:H-g}.

To prove $d \omega=\iota^* H$, with $\omega$ given in equation~\eqref{eq:omega}, we choose the conjugacy class representative~\eqref{eq:representative}, which satisfies equation~\eqref{eq:brane} with constant $u$.
With the one-form $\theta$ of equation~\eqref{eq:brane-currents}, one has
\begin{equation}\label{eq:app-gdg}
g^{-1}dg=\Ad_s\big(\Ad_u^{-1}-1\big)\theta\,,\qquad
\Ad_{wg}=\Ad_s\,\Ad_u\,\Ad_s^{-1}\,.
\end{equation}
Inserting these expressions into the definition of $\omega$ in equation~\eqref{eq:omega}, we use the identity
\begin{equation}\label{eq:app-operator}
\frac{1+\Ad_u}{1-\Ad_u}\,\big(\Ad_u^{-1}-1\big)=1+\Ad_u^{-1}
\end{equation}
together with the invariance and cyclicity of the supertrace, and find
\begin{equation}\label{eq:app-omega}
\omega=\frac14\,\str\Big(\big(\Ad_u^{-1}\theta-\theta\big)\wedge
\big(\Ad_u^{-1}\theta+\theta\big)\Big)-i\,\str\big(j_1\wedge j_3\big)
=\frac12\,\str\big(\theta\wedge\Ad_u\theta\big)-i\,\str\big(j_1\wedge j_3\big)\,.
\end{equation}
Since $u$ is constant and $\theta$ obeys the Maurer--Cartan equation, the exterior derivative of this expression is
\begin{equation}\label{eq:app-domega}
d\omega=\frac12\,\str\big(-\theta\wedge\theta\wedge\Ad_u\theta
+\theta\wedge\Ad_u\theta\wedge\Ad_u\theta\big)-i\,d\,\str\big(j_1\wedge j_3\big)\,.
\end{equation}
On the other hand, expanding the cube of $g^{-1}dg$ gives
\begin{equation}\label{eq:app-cubic-brane}
\str\big((g^{-1}dg)^3\big)=\str\Big(\big(\Ad_u^{-1}\theta-\theta\big)^3\Big)
=3\,\str\big(-\theta\wedge\theta\wedge\Ad_u\theta
+\theta\wedge\Ad_u\theta\wedge\Ad_u\theta\big)\,,
\end{equation}
where the two pure cubic terms cancel by the invariance of the supertrace. Comparing with the expression for $H$ in equation~\eqref{eq:H-g}, evaluated for the representative~\eqref{eq:representative}, we conclude that $\omega$ is a potential for the pullback of $H$, as stated in equation~\eqref{eq:omega}.

Moreover, if we restrict the potential to the brane and use the currents in equation~\eqref{eq:brane-currents}, the fermionic currents along the brane are
\begin{equation}\label{eq:app-fermionic-brane}
j_1=\tfrac12\big(\Ad_u^{-1}\theta^F+i\,\theta^F\big)\,,\qquad
j_3=\tfrac12\big(\Ad_u^{-1}\theta^F-i\,\theta^F\big)\,.
\end{equation}
The squares of $\Ad_u^{-1}\theta^F$ and $\theta^F$ have vanishing supertrace, and the two mixed terms are equal, so we find
\begin{equation}\label{eq:app-rr}
\str\big(j_1\wedge j_3\big)=-\frac{i}{2}\,\str\big(\theta^F\wedge\Ad_u\theta^F\big)\,.
\end{equation}
Inserting this expression in the potential~\eqref{eq:app-omega}, we observe that the fermionic terms cancel, and we obtain
\begin{equation}
\omega=\frac12\,\str\big(\theta^B\wedge\Ad_u\theta^B\big)\,.
\end{equation}
This proves the expression in equation~\eqref{eq:omega-brane}.

\section{Equations of motion of the D3-brane}\label{app:brane-eom}

In this appendix we derive the equations of motion~\eqref{eq:brane-eom} of the AdS$_2\times$S$^2$ brane from the action in equation~\eqref{eq:D3-action}. We use the coordinates of equation~\eqref{eq:slicing} and the orientations
\begin{equation}\label{eq:appB-orientation}
\begin{gathered}
v_A=\cosh\rho\,dt\wedge d\rho\,,\qquad
v_S=\sin\beta\,d\beta\wedge d\gamma\,,\\
\mathrm{vol}_{\AdS_3}=\cosh^2\!\psi\,d\psi\wedge v_A\,,\qquad
\mathrm{vol}_{\Sph^3}=-\sin^2\!\alpha\,d\alpha\wedge v_S\,,
\end{gathered}
\end{equation}
which are induced by the sigma model: with the parametrisation~\eqref{eq:brane-coordinates}, the bosonic part of the three-form~\eqref{eq:H} equals $2(\mathrm{vol}_{\AdS_3}+\mathrm{vol}_{\Sph^3})$, in agreement with the background~\eqref{eq:nh-background}. With these orientations, the fields of the pure NSNS brane take the values in equation~\eqref{eq:fields-NSNS}. In the background of equation~\eqref{eq:nh-background}, the potentials can be chosen as
\begin{equation}\label{eq:appB-potentials}
\begin{aligned}
B&=2R^2\chi\,\big(\mu_A\,v_A-\mu_S\,v_S\big)\,,&\qquad
\mu_A&=\tfrac12\big(\psi+\sinh\psi\cosh\psi\big)\,,\\[2mm]
g_sC_2&=2R^2\kappa\,\big(\mu_A\,v_A-\mu_S\,v_S\big)\,,&\qquad
\mu_S&=\tfrac12\big(\alpha-\sin\alpha\cos\alpha\big)\,,\\[2mm]
C_4&=\frac{4R^4\chi\kappa}{g_s}\,\mu_A\mu_S\,v_A\wedge v_S\,.
\end{aligned}
\end{equation}
They reproduce $\text{H}$ and $F_3$, and the five-form field strength that is invariant under the gauge transformations of the Wess--Zumino coupling vanishes,
\begin{equation}\label{eq:appB-F5}
F_5=dC_4+\text{H}\wedge C_2=0\,.
\end{equation}
The brane sits at fixed $\psi$ and $\alpha$, and for constant fields its worldvolume field takes the form
\begin{equation}\label{eq:appB-F}
\mathcal{F}=R^2\big(\hat{\mathsf{e}}\,v_A+\hat{\mathsf{b}}\,v_S\big)\,,\qquad
\hat{\mathsf{e}}=2\chi\mu_A+\phi_A=\cosh^2\!\psi\;\mathsf{e}\,,\qquad
\hat{\mathsf{b}}=-2\chi\mu_S+\phi_S=\sin^2\!\alpha\;\mathsf{b}\,,
\end{equation}
where the constants $\phi_A$ and $\phi_S$ are the components of $2\pi\alpha' f/R^2$. The last expressions follow from the definition of $\mathsf{e}$ and $\mathsf{b}$ in equation~\eqref{eq:worldvolume-field} as the components of $\mathcal{F}$ in orthonormal frames. According to equation~\eqref{eq:slicing}, the induced metric describes AdS$_2$ and S$^2$ with radii $R\cosh\psi$ and $R\sin\alpha$. If we orient the frames along $v_A$ and $v_S$, their area forms are therefore $R^2\cosh^2\!\psi\,v_A$ and $R^2\sin^2\!\alpha\,v_S$.

In the action~\eqref{eq:D3-action}, the Dirac--Born--Infeld term therefore reduces to the Lagrangian in equation~\eqref{eq:DBI}, multiplied by the induced volume form $R^4\cosh^2\!\psi\,\sin^2\!\alpha\;v_A\wedge v_S$. With the potentials in equation~\eqref{eq:appB-potentials}, the brane action becomes
\begin{equation}\label{eq:appB-action}
S_{\mathrm{D3}}
=\frac{T_3R^4}{g_s}\int v_A\wedge v_S\,
\mathcal L_{\mathrm{D3}}\,,\qquad
\mathcal L_{\mathrm{D3}}=\cosh^2\!\psi\,\sin^2\!\alpha\;L(\mathsf{e},\mathsf{b})
-2\kappa\big(\mu_A\hat{\mathsf{b}}-\mu_S\hat{\mathsf{e}}
+2\chi\mu_A\mu_S\big)\,,
\end{equation}
where the term proportional to $\chi\mu_A\mu_S$ is the contribution of $C_4$. Since the induced metric contains the derivatives of $\psi$ and $\alpha$ only quadratically, and since the potentials have no components along $d\psi$ and $d\alpha$, the equations of motion of the embedding reduce to the derivatives of $\mathcal L_{\mathrm{D3}}$ with respect to $\psi$ and $\alpha$ at fixed gauge field, i.e.\ at fixed $\phi_A$ and $\phi_S$. The equations of motion of the gauge field hold, since the pullbacks of $\text{H}$ and $F_3$ to the brane vanish and the Dirac--Born--Infeld current is a constant combination of $v_A$ and $v_S$. In the derivatives we use the identities
\begin{equation}\label{eq:appB-identities}
L-\mathsf{e}\,\mathsf{d}=-\varepsilon\,,\qquad
L+\mathsf{b}\,\mathsf{h}=-\varepsilon^{-1}\,,\qquad
\varepsilon=\frac{\sqrt{1+\mathsf{b}^2}}{\sqrt{1-\mathsf{e}^2}}\,,
\end{equation}
which follow from equation~\eqref{eq:dual-fields}. Here $\varepsilon$ is the energy density of the brane per unit induced volume, in the units of equation~\eqref{eq:DBI}.

The derivatives of $\mathcal L_{\mathrm{D3}}$ receive three types of contributions. The variation of the induced metric enters through the prefactor of $L$ and through the relations between $(\mathsf{e},\mathsf{b})$ and $(\hat{\mathsf{e}},\hat{\mathsf{b}})$. The pullback of $B$ shifts the field strength by the contraction of $\text{H}$ with the variation,
\begin{equation}\label{eq:appB-shifts}
\partial_\psi\hat{\mathsf{e}}=2\chi\cosh^2\!\psi\,,\qquad
\partial_\alpha\hat{\mathsf{b}}=-2\chi\sin^2\!\alpha\,,
\end{equation}
and the Wess--Zumino term contributes through $\hat{\mathsf{e}}$ and
$\hat{\mathsf{b}}$. Altogether, we find
\begin{equation}\label{eq:appB-derivatives}
\begin{aligned}
0=\partial_\psi\mathcal L_{\mathrm{D3}}&=
\underbrace{2\sinh\psi\cosh\psi\sin^2\!\alpha\,\big(L-\mathsf{e}\,\mathsf{d}\big)}_{\text{induced metric}}
+\underbrace{\sin^2\!\alpha\;\mathsf{d}\;\partial_\psi\hat{\mathsf{e}}}_{\text{shift of }\mathcal{F}}
-\underbrace{2\kappa\cosh^2\!\psi\;\hat{\mathsf{b}}}_{\text{Wess--Zumino}}\,,\\[4mm]
0=\partial_\alpha\mathcal L_{\mathrm{D3}}&=
\underbrace{2\sin\alpha\cos\alpha\cosh^2\!\psi\,\big(L+\mathsf{b}\,\mathsf{h}\big)}_{\text{induced metric}}
-\underbrace{\cosh^2\!\psi\;\mathsf{h}\;\partial_\alpha\hat{\mathsf{b}}}_{\text{shift of }\mathcal{F}}
+\underbrace{2\kappa\sin^2\!\alpha\;\hat{\mathsf{e}}}_{\text{Wess--Zumino}}\,.
\end{aligned}
\end{equation}
In the Wess--Zumino terms, the contribution of $C_4$ has cancelled the terms that depend on the gauge of $C_2$, and what remains is the contraction of $F_3$ with the variation, wedged with $\mathcal{F}$. Inserting the shifts in equation~\eqref{eq:appB-shifts} and the identities in equation~\eqref{eq:appB-identities}, the equations of motion become
\begin{equation}\label{eq:appB-force-balance}
\varepsilon\tanh\psi=\chi\,\mathsf{d}-\kappa\,\mathsf{b}\,,\qquad
\varepsilon^{-1}\cot\alpha=\chi\,\mathsf{h}+\kappa\,\mathsf{e}\,.
\end{equation}
The first equation balances the tension of the brane against the forces that the NSNS and RR fluxes exert on the dissolved fundamental strings and D1-branes. Dividing it by $\varepsilon$, multiplying the second equation by $\varepsilon$ and using equation~\eqref{eq:dual-fields}, we obtain the equations of motion~\eqref{eq:brane-eom}. If we had chosen the opposite orientation of the brane, only the sign of $\kappa$ would have been changed.

\bibliographystyle{JHEP}
\bibliography{references}

\end{document}